\newif\iftechreport

\techreporttrue

\iftechreport
\documentclass[10pt,technote,comsoc]{IEEEtran}
\else
\documentclass[10pt,journal,compsoc]{IEEEtran}
\fi
\iftechreport
\else
	\ifCLASSOPTIONcompsoc
	  \usepackage[nocompress]{cite}
	\else
	  \usepackage{cite}
	\fi
\fi

\usepackage{amsmath,amssymb,amsfonts}
\usepackage{algorithmic}
\usepackage{graphicx}
\usepackage{url}
\usepackage{subfigure} 
\usepackage{listings} 
\usepackage{xcolor} 
\usepackage{bm}		

\begin{document}
\bstctlcite{IEEEexample:BSTcontrol}
%
\iftechreport
\title{Technical Report: Edge-centric Programming for IoT Applications with EdgeProg}
\else
\title{Edge-centric Programming for IoT Applications with Automatic Code Partitioning}
\fi
%
%
%
%

\iftechreport
\author{Borui Li, and, Wei Dong}
\else
\author{Borui~Li,~\IEEEmembership{Student Member,~IEEE,}
        and~Wei~Dong,~\IEEEmembership{Member,~IEEE,}
\thanks{B. Li and W. Dong are with the College of Computer Science, Zhejiang University, China, and, Alibaba-Zhejiang University Joint Institute of Frontier Technologies, E-mail: \{\textit{libr, dongw}\}@emnets.org. Wei Dong is the corresponding author.}
}
\fi

%
%

\iftechreport\else
\markboth{IEEE TRANSACTIONS ON COMPUTERS, VOL. 32, NO. 3, MARCH 2021}%
{Li \MakeLowercase{\textit{et al.}}: EdgeProg: Edge-centric Programming for IoT Applications}
\fi
%



\iftechreport
\else
\IEEEtitleabstractindextext{%
\begin{abstract}
IoT application development usually involves separate programming at the device side and server side. 
While separate programming style is sufﬁcient for many simple applications, it is not suitable for many complex applications that involve complex interactions and intensive data processing.

We propose \textit{EdgeProg}, an edge-centric programming approach to simplify IoT application programming, motivated by the increasing popularity of edge computing. 
With EdgeProg, users could write application logic in a centralized manner with an augmented If-This-Then-That (IFTTT) syntax and virtual sensor mechanism. 
The program can be processed at the edge server, which can automatically generate the actual application code and intelligently partition the code into device code and server code, for achieving the optimal latency. 
EdgeProg employs dynamic linking and loading to deploy the device code on a variety of IoT devices, which do not run any application-speciﬁc codes at the start. 
Results show that EdgeProg achieves an average reduction of 20.96\%, 27.8\% and 79.41\% in terms of execution latency, energy consumption, and lines of code compared with state-of-the-art approaches.
\end{abstract}
\fi
\iftechreport
\else
}
\fi

\maketitle

\iftechreport\else
\IEEEdisplaynontitleabstractindextext
\fi

%
\iftechreport\else
\IEEEpeerreviewmaketitle
\fi

\iftechreport
\section{Introduction}\label{sec:introduction}
Internet of Things (IoT) application development usually involves separate programming at the device side and server side.
\else
\IEEEraisesectionheading{
	\section{Introduction}\label{sec:introduction}
}
\IEEEPARstart{I}{nternet} of Things (IoT) application development usually involves separate programming at the device side and server side.
\fi
For example, consider a smart plant application. 
Users can program an IoT node like Arduino to sense the soil humidity of a plant. 
The sensing data can then be transmitted to the back-end server for further analysis. 

This separate programming style is sufficient for many simple applications. 
However, it is not suitable for many complex applications that involve complex interactions and intensive data processing.

\textbf{Complex interactions. }
Consider the following application: a user wants to turn on an LED when a sensor attached to a door detects an open event.
With the traditional programming style, the application logic would be scattered among different sensor nodes.
Developers should cope with the complex data stream and network interactions between sensor nodes, which leads to increased system complexity and reduced manageability. 

\textbf{Intensive data processing. }
Consider a speech recognition application. 
A simple way of designing such a system would deliver all the sensor data to the server running the sophisticated recognition algorithm. 
This approach may consume excessive energy due to a large number of transmissions. 
A different approach is to run the recognition algorithm on the IoT device. 
This approach, however, may cause excessive delays due to the insufficient computation power of the device.
Separate programming requires the programmer to make proper decisions, which is quite difficult.

We advocate here a different programming approach, motivated by the increasing popularity of edge computing.
In the edge computing paradigm, a number of IoT nodes can perform sensing and actuation.
These nodes are connected to a local edge that can perform sophisticated computation. 
Moreover, edge servers usually have power supplies and are less constrained by energy. Edge computing can offer low processing delay and better privacy. 

Taking advantage of the edges, we have developed EdgeProg---a new programming style and software architecture to greatly simplify IoT application programming, resulting in a generic IoT system that can be reprogrammed for a variety of applications without significant loss of overall system efficiency. 

To use EdgeProg, developers write a program in a high-level language integrating the whole application logic of an IoT application. 
This program can further be processed at the edge server, which can automatically generate the actual application code and intelligently partition the code into device code and server code. 
We call this approach \textit{edge-centric} since developers can regard the program as if it runs on the edge.
More importantly, ordinary IoT nodes do not run any application-specific codes at the start. 
When the program is first executed, the device code will be automatically loaded onto the memory of IoT nodes.
Nevertheless, this edge-centric programming process raises some challenges:
\begin{itemize}
	\item How to design an edge-centric language that could support multi-device interaction and data-intensive computation?
	\item How to partition the user-perceived program to achieve the best delay performance or save most energy?
	\item How to design a mechanism so that heterogeneous sensor nodes can dynamically load the device-side code and execute it in an efficient manner?
\end{itemize}

In order to support edge-centric programming and speed-up the application development process, we design a coherent language for specifying the multi-device interaction based on the widely-adopted programming model, IFTTT (IF-This-Then-That)~\cite{ur2016trigger}.
To further enhance the expressiveness and adopt the data-intensive computation, we extend the traditional IFTTT syntax with the virtual sensor, which accelerates developers to design their own data processing logic with machine learning techniques.

EdgeProg conducts automatic code partitioning which fully leverages the computation ability of each device and achieves optimal end-to-end latency.
We abstract the user-written program as a data flow graph, formulate the partitioning problem as an integer programming (ILP) problem and leverage the efficient solver \texttt{lp\_solve} to obtain the optimal partition.

We implement EdgeProg with Contiki OS for its cross-platform support and the ability to load the optimized executable at runtime with dynamic linking and loading technique.
An alternative approach to change the application logic during its execution is exploiting virtual machines (VMs) or using a scripting language.
Nevertheless, we do not adopt the alternatives due to they introduce considerable overhead than dynamic linking and loading.

We implement EdgeProg and evaluate its performance extensively.
Results show that:
(1) EdgeProg programming language can express diverse IoT application logic and reduces the lines of code needed by 79.41\% on average.
(2) For the execution time of generated applications, EdgeProg achieves a 20.96\% reduction on average, and up to 99.05\% latency reduction across the five real-world applications under all settings compared with state-of-the-art partitioning systems such as Wishbone~\cite{newton2009wishbone} and RT-IFTTT~\cite{heo2017rt}. 
(3) The partitioned application generated by EdgeProg saves 14.8\% and 40.8\% energy on average compared to Wishbone and RT-IFTTT.
(4) For application run-time, the dynamic linking and loading technique outperforms than design alternatives such as virtual machine (by 9.98$\times$) and scripting languages (by 6.37$\times$).
(5) The profiling methods adopted by EdgeProg achieve 90\%+ and 85\%+ accuracy for over 98\% test cases.
The contributions of this work are summarized as below:
\begin{itemize}
	\item We present EdgeProg, an \textit{edge-centric} programming system for IoT applications.
	The EdgeProg language relieves developers from scattered application logic and enables them to express their logic in an easy-to-use way.
	\item We formulate the code partitioning problem as an ILP problem to minimize the makespan of the task or the energy consumption.
	The partitioning algorithm optimizes the placement of each stage in an application with consideration of both processing and network cost.
	\item We implement EdgeProg and evaluate EdgeProg massively with real-world applications and benchmarks.
	Results show that EdgeProg achieves better latency reduction compared with state-of-the-art approaches and fewer lines of code.
\end{itemize}

Compared to the conference version of EdgeProg~\cite{li2020edgeprog}, this journal version contains the following important extensions.

\begin{itemize}
	\item We add more detailed descriptions in Section~\ref{sec:background-and-usage} to better illustrate the usage of EdgeProg.
	\item Besides latency, we facilitate EdgeProg with the ability to optimize the energy consumption of the edge-device integrated system. Based on the efforts, we reconstruct Section~\ref{sec:code-partitioning} with the new analytical formulation of the energy optimization problem and its solution. Also, we present descriptions of the building blocks towards energy optimization in Section~\ref{sec:overview}.
	\item We add more details about how EdgeProg generates the binaries with Contiki OS in Section~\ref{sec:executable-generator}.
	\item We present a much more detailed evaluation on EdgeProg, especially on the energy optimization performance, in Section~\ref{sec:evaluation}.
	\item We add Section~\ref{sec:discussion} to discuss several important issues about EdgeProg.
\end{itemize}

The rest of this article is structured as follows.
Section~\ref{sec:background-and-usage} describes the background and usage of EdgeProg.
Section~\ref{sec:overview} overviews the design goals and building blocks.
Section~\ref{sec:design} presents the design details.
Section~\ref{sec:evaluation} shows the evaluation results.
Section~\ref{sec:discussion} discusses several important issues about EdgeProg and Section~\ref{sec:related-work} introduces the related work.
Finally, Section~\ref{sec:conclusion} concludes the paper.

\section{Background and EdgeProg Usage}
\label{sec:background-and-usage}
In this section, we briefly introduce the background of the dynamic linking and loading technique of IoT devices  used in EdgeProg.
Then we present the usage of EdgeProg with a simple smart home application.

\subsection{Dynamic linking and loading of IoT devices}
\label{sec:background}
Dynamic linking and loading is one of the over-the-air reprogramming techniques for IoT devices.
As its name suggests, reprogramming with dynamic linking and loading technique owns a linking phase and a loading phase.
In the linking phase, the on-device reprogrammer first parses the structured information of a file in standard executable and linkable format (ELF) or its variants (e.g., CELF~\cite{dunkels2006run} and SELF~\cite{dong2009dynamic}).
Then the reprogrammer allocates ROM and RAM for the data and text segment in the ELF file and performs relocation.
The relocation is to patch the data and text segment with real in-memory addresses of the symbols, which are found in the symbol table or calculated using the relocation information in the ELF.
Once the linking phase is complete, the reprogrammer writes text segments to the allocated ROM and copies data segments to the RAM, which is called the loading phase.
So far, the binary is loaded and ready to be executed.

Compared with the alternatives such as virtual machine~\cite{levis2002mate,dunkels2006run,reijers2018capevm} and bootloader~\cite{dong2010elon}, dynamic linking and loading obtains several inherent merits.
(1) High long-term efficiency because it runs native code rather than virtual machine code.
(2) Reboot-less update, which is also energy-saving.
The recent container technology is also a potential alternative to dynamic linking and loading.
However, it heavily relies on Linux services such as \texttt{cgroup} and \texttt{namespace}, which is not available in the resource-constrained IoT devices such as Arduino, TelosB and STM32.


\subsection{EdgeProg Usage}
\label{sec:motivating-examples}
\begin{figure}[tbp]
	\centering
	\subfigure[Pipeline illustration of \texttt{SmartHomeEnv} application.]{
		\begin{minipage}[t]{\linewidth}
			\label{fig:motivating_example_1}
			\centering
			\includegraphics[width=.9\linewidth]{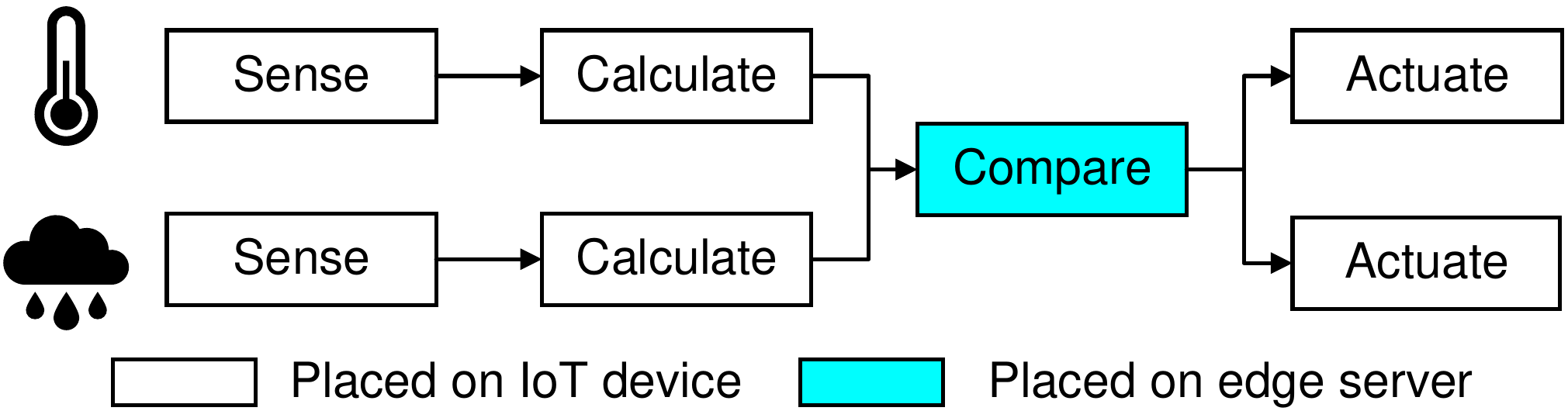}
			\vspace{0.5em}
		\end{minipage}%
	}%

	\subfigure[Pipeline illustration of \texttt{SmartDoor} application.]{
		\begin{minipage}[t]{\linewidth}
			\centering
			\label{fig:motivating_example_2}
			\includegraphics[width=.95\linewidth]{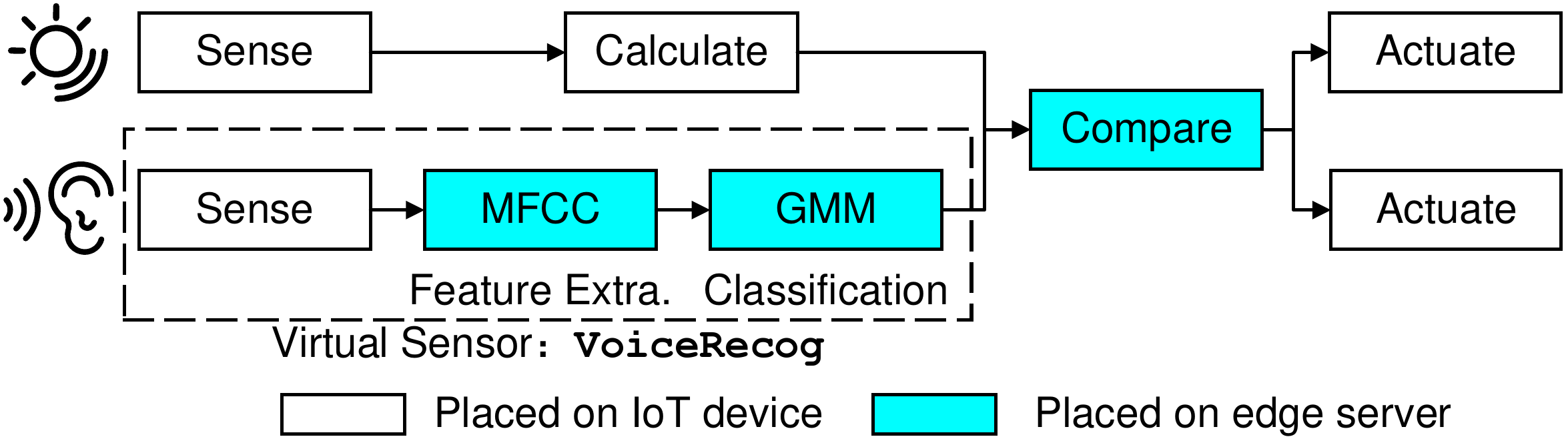}
			\vspace{0.5em}
		\end{minipage}
	}%
	\centering
	\vspace{-0.5em}
	\caption{Illustration of execution stages of the two examples. Each block represents one stage, and data exchange occurs between consecutive stages.}
	\label{fig:motivating_example}
\end{figure}

We excerpt a simple smart home project named \texttt{SmartHomeEnv} from \url{smarthome.com} to illustrate how EdgeProg can be used. 
\texttt{SmartHomeEnv} takes the temperature and humidity data from two IoT nodes as input, turns on the air conditioner and dryer if the two readings exceed fixed thresholds, as shown in Fig.~\ref{fig:motivating_example_1}.
The two nodes are wirelessly connected to an edge server, which could be a PC or other devices that own strong computing ability.

In the traditional approach, two sensors are pre-installed with an application-specific code with functions like periodically transmitting sensor values to the edge server. 
The edge server further processes these readings and interacts with the sensors with pre-defined interfaces.

With EdgeProg, in contrast, the two sensors are pre-installed with an “idle” program without any application-specific logic. 
The whole application logic is expressed in an enhanced IFTTT-like language, which is interpreted and processed at the edge server. 
Fig.~\ref{fig:motivating-example-code} shows an EdgeProg application of \texttt{SmartHomeEnv}. 
Lines 2-4 describe the devices (\texttt{A}, \texttt{B}, \texttt{E}) and their interfaces (e.g., the \texttt{HUMIDITY} of device \texttt{B}) used in this application with keyword \texttt{Configuration}.
With the information above, lines 5-6 specify the application logic following the IFTTT manner.
The edge server automatically partitions the codes into two components, i.e., device-side components and edge-side components. 
The former are compiled to a loadable module and dispatched to the sensor nodes. 
Once notified, the “idle” program in the IoT node can dynamically load application-specific module for execution via dynamic linking and loading technique. 

EdgeProg also supports building complex applications targeting intensive data processing with the virtual sensor.
Fig.~\ref{fig:motivating_example_2} illustrates a \texttt{SmartDoor} application with several steps for voice recognition.
With EdgeProg, the latter steps could be expressed by a virtual sensor \texttt{VoiceRecog}.
Opposite to the physical or hardware sensor, a virtual sensor is a logical entity that abstracts the data sensed by real sensors which could be located at different places, which we will further describe in Section~\ref{sec:programming-language}.

A key feature of EdgeProg is that it can automatically partition the whole application code to optimize the execution performance, which is increasingly essential for computation-intensive IoT tasks such as speech recognition and video surveillance.
As illustrated in both figures of Fig.~\ref{fig:motivating_example}, EdgeProg places each stage on appropriate devices for best performance (i.e., execution time or energy consumption).
The MFCC and GMM algorithms in Fig.~\ref{fig:motivating_example_2} may be too heavyweight for resource-constrained IoT devices such as TelosB or Arduino.
Hence, EdgeProg will automatically partition this task to the edge-side if it yields better performance than placing it on the device.

\begin{figure}
	\centering
	\begin{lstlisting}
Application SmartHomeEnv{
	Configuration{TelosB A(TEMPERATURE);
		TelosB B(HUMIDITY);
		Edge E(turnOnAC, turnOnDryer);}
	Rule{IF (A.TEMPERATURE > 30 && B.HUMIDITY > 70)
		THEN (E.turnOnAC && E.turnOnDryer)}
}
	\end{lstlisting}
	\vspace{-1em}
	\caption{Code snippets of \texttt{SmartHomeEnv} application.}
	\label{fig:motivating-example-code}
	\vspace{-1em}
\end{figure}

\section{EdgeProg Overview}
\label{sec:overview}
In this section, we first discuss the design goals of EdgeProg, overview our system design, and introduce some essential components.

\subsection{Design Goals}
\label{sec:design-goals}
\begin{itemize}
	\item \textbf{Edge-centric.} Compared to the traditional scattered programming manner, EdgeProg should provide users with an edge-centric approach to create the application, which indicates that users need not to break down the application logic into pieces during development.
	\item \textbf{Cost-aware.} The timeliness and energy consumption are recognized as critical costs of an edge-device coordinated application. The ability to deliver a cost-optimal solution of a given input is one of the requirements in EdgeProg's design.
	\item \textbf{Automatic.} By automatic, we mean that EdgeProg should conceive details which have no benefit for users to express their ideas and removes human from the loop to simplify and accelerate the application development. 
\end{itemize}

\subsection{EdgeProg Architecture}
\label{sec:architecture}
\begin{figure*}[tbp]
	\centering
	\includegraphics[width=.85\linewidth]{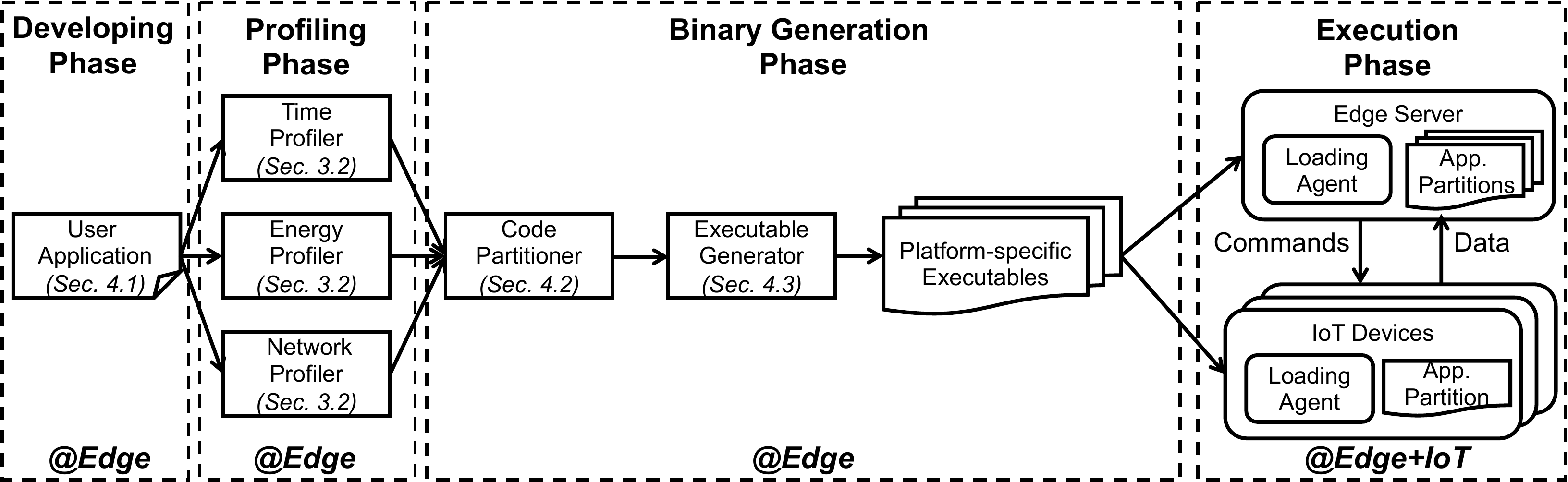}
	\vspace{-1em}
	\caption{System overview of EdgeProg.}
	\label{fig:overview}
	\vspace{-1em}
\end{figure*}

In Fig.~\ref{fig:overview}, we show a birds-eye view of EdgeProg's system architecture and functional workflow, considering a developing phase, a profiling phase, a binary generation phase, and an execution phase.
Users can directly write the application code in an edge-centric manner, i.e., without following the distributed programming style or considering the physical placement of each stage (see Section~\ref{sec:programming-language} for details).
The system takes the user code as input, preprocesses and feeds it into the \textit{code partitioner}.
With the help of the \textit{time, energy and network profile} of each device or application, the code partitioner finds the optimal partition and placement of each stage using our partitioning algorithm.
Processed by the \textit{code generator}, the user-written code is then transformed into the compilable code and compiled to executable or loadable binary by the \textit{code compiler}.
Finally, the executables are disseminated to the devices over the air or deployed on the edge device if necessary.

\textbf{User Input.}
The user input is written in the EdgeProg programming model, which centers around the notion of \texttt{Rules} that specifies the application logic with sensor data, actuator presented by the devices, specified by \texttt{Interfaces}, or virtual sensors' output, specified by \texttt{Implementation}.
Detailed features of the EdgeProg programming language are specified in Section~\ref{sec:programming-language}.

\textbf{Code Partitioner.}
The code partitioner is responsible for generating the optimal partition of the user-input applications.
We will further give a detailed description in Section~\ref{sec:code-partitioning}.

\textbf{Time Profiler.}
The execution time of each stage on different devices represents the different computation capabilities of each device, which is one of the critical inputs to the code partitioner.
Similar to~\cite{newton2009wishbone,cuervo2010maui}, EdgeProg leverages a profiling phase to obtain the execution time on different platforms.
For the low-end sensor nodes, we exploit the cycle-accurate simulators such as \texttt{MSPsim} for MSP430-based nodes (\textit{e.g.,} TelosB) and \texttt{Avrora} for AVR-based nodes (\textit{e.g.,} MicaZ) to get the timing information.
For the high-end devices such as Raspberry Pi, profiling it with simulators will be less accurate than the low-end ones mainly due to these powerful devices employ automatic frequency scaling strategy, which reduces the accuracy of a simulator.
However, executing on the real device and collect raw timing data is painful and sometimes infeasible due to the hardware interface limit of edge servers.
Hence, we choose a near cycle-accurate simulator named \texttt{gem5} for profiling high-end devices.
We will evaluate the profiling accuracy in Section~\ref{sec:profiling-accuracy}.

\textbf{Energy Profiler.}
In EdgeProg, an energy profiler is necessary when the objective is to minimize energy consumption.
Hence, we build and maintain an energy profile of each device.
To be more specific, the profile mainly contains the power under idle state, productive state and network TX/RX.
We adopt an automated hardware knowledge based generation approach based on weak supervision learning~\cite{hsiao2019automating,hsiao2020creating} to generate the energy profile of each device.
This learning-based approach reduces the instability and randomness brought by human error, which is more scalable when generating profiles for new devices.

\textbf{Network Profiler.}
Network condition (\textit{e.g.,} bandwidth) is also a critical metric being fed into the partitioner.
In order to predict the network condition when the application is deployed, we leverage the multiple-output support vector regression (M-SVR) algorithm~\cite{sanchez2004svm} since it generates a series of prediction results representing the future network condition in a sequence of intervals.
In our temporary implementation, the network profiler contains the prediction of the WiFi and Zigbee network.
Raw observations such as the bandwidth and received signal strength indicator (RSSI), which is sampled by the loading agent every 60s in order not to influence the regular network transmission, are fed into the M-SVR.
Furthermore, when the IoT device is deployed with an application and periodically uploads the data or receiving the commands, the network profiler piggybacks the measurement data with the regular sensor data or commands to further reduce the energy overhead.
The predictor outputs the future throughput estimation and per-packet transmission time for further fine-grained time calculation in Section~\ref{sec:code-partitioning}.
Here, since the predicting algorithm acts as a black-box in our system, EdgeProg can use other prediction models instead of the M-SVR model.

\textbf{Code Generator.}
The generated optimal partition is processed by the code generator to translate the high-level EdgeProg code into the compilable C code, detailed in Section~\ref{sec:executable-generator}.

\textbf{Code Compiler.}
Fed by the compilable code, the code compiler generates the executables for the target platform and starts dissemination.
In our current implementation, EdgeProg supports four MCU architectures (ATmega, MSP, ARM and x86) with four platforms.

\textbf{Loading Agent.}
At the very beginning of our system, there is no application-specific logic running on the node except a loading agent.
The loading agent periodically communicates with the edge server for new loadable applications.
Once the application is compiled by the compiler and starts dissemination, the loading agent on the deployment destination detects, verifies and receives the executable and dynamically runs it.
Moreover, using the wireless channel to dispatch the applications may be unstable due to the existence of wireless interference.
Hence, we also advocate a wired loading agent to support disseminating the binaries through USB (for TelosB) and Ethernet (for Raspberry Pi).

\section{System Design}
\label{sec:design}
In this section, we will first present the design of EdgeProg programming language and highlight the features which enable integrated development. 
Then we will describe how EdgeProg obtains the optimal partition of the input IoT application with full awareness of the user-perceived event handling latency or energy consumption, including details about the problem formulation and its solution algorithm.
Finally, we demonstrate how EdgeProg generates the application code to be disseminated to both devices and edge servers.

\subsection{EdgeProg Programming Language}
\label{sec:programming-language}

In order to tackle the problem of existing scattered programming style and accelerate the application development process, EdgeProg adopts a rule-based domain-specific language (DSL) for developers to build their applications. 
An EdgeProg application is typically organized as three parts: \textit{configuration}, \textit{implementation} and \textit{rule}.
As shown in Fig.~\ref{fig:programming-language}, we use the SmartDoor application described in Section~\ref{sec:motivating-examples} as an example to illustrate three critical features in the following.
\iftechreport
More examples could be found in Appendix~\ref{sec:appendix-language}.
\else
More examples could be found in~\cite{li2021edgeprogtechreport}.
\fi

\begin{figure}
	\centering
	\begin{lstlisting}
Application SmartDoor{
	Configuration{
		RPI A(MIC, DOOR_UNLOCK, OPEN_DOOR);
		TelosB B(LIGHT_SOLAR); (*@\label{loc:programming-language-telosb-library} @*)
	}
	Implementation{
		VSensor VoiceRecog("FE, ID"){
			VoiceRecog.setInput(A.MIC);
			FE.setModel("MFCC");
			ID.setModel("GMM", "open.gmm"); (*@\label{loc:programming-language-vsensor-library} @*)
			VoiceRecog.setOutput(<string_t>,"open");
		}
	}
	Rule{
		IF(VoiceRecog=="open" && B.LIGHT_SOLAR<100)
		THEN(A.DOOR_UNLOCK && A.OPEN_DOOR)
	}
}
	\end{lstlisting}
	\vspace{-1em}
	\caption{Code snippets of the SmartDoor application.}
	\label{fig:programming-language}
	\vspace{-1em}
\end{figure}

\textbf{Edge-centric programming model.}
In order to achieve the edge-centric design goals of EdgeProg, our programming model should focus users more upon the global behavior other than implementation details.
Hence, EdgeProg enables developers to organize their application centered with the overall application logic using keyword \texttt{Rule}.
There exist several DSLs enabling developers to focus on upper logic, as known as the macro-programming model, in sensornet researches such as Kairos~\cite{gummadi2005macro} and Regiment~\cite{newton2007regiment}.
Nevertheless, existing works fall flat nowadays due to the constraint on application portability or lack of actuation.
IFTTT programming shows its simpleness and effectiveness in existing researches~\cite{ur2014practical,heo2017rt,huang2015supporting} when expressing the high-level application logic, and this programming approach is widely adopted in state-of-the-art industrial solutions such as Samsung SmartThings and Microsoft Flow.
By early 2017, the website \url{ifttt.com} had gathered over 320,000 IFTTT programs~\cite{ur2016trigger} and the numbers are still increasing dramatically.
Therefore, we leverage an IFTTT-like grammar for enabling users to express their idea in a unified and explicit manner, as illustrated in lines 14-17 of Fig.~\ref{fig:programming-language}.
Moreover, we augment the IFTTT grammar with \texttt{Configuration} and \texttt{Implementation} to make users express the detailed definition and specification of necessary components used in the \texttt{Rule} part.

\textbf{Full support of virtual sensor.}
In order to accommodate the intensive data processing in the nowaday IoT scenario, we enhance our DSL with the \textit{virtual sensor}.
Traditional hardware sensors generally produce raw measurements of physical properties such as the moisture value or light intensity, which are unprofitable unless being transformed into high-level domain-dependent information.
Furthermore, capturing valuable information usually requires the coordination of multiple hardware sensors, e.g., detecting fire hazards with both temperature and smoke sensor.
In order to tackle the limitations above and make sensor data processing more flexible, existing works combine the readings of multiple sensors for event detection.
For example, SenseHAR~\cite{jeyakumar2019sensehar} advocates an activity recognition system that abstracts the data of several inertial sensors from different devices using a sensor fusion network.
Similarly, LiKamWa \textit{et al.}\cite{likamwa2013moodscope} measure the user's mental state based on the interactions with the smartphone.
Virtual sensors act as a black-box providing the indirect measurements or events, which are typically physically immeasurable, by combining sensed data from several hardware sensors with data processing algorithms.
EdgeProg embraces this technique as one of the extensions to standard IFTTT syntax to provide easy-to-use yet expressive handling for intensive data processing.

As shown in Fig.~\ref{fig:programming-language}, lines 4-12 list the configuration of a virtual sensor, \texttt{VoiceRecog}, to recognize whether the input voice fragment produced by interface \texttt{A.MIC} stands for "open" or not.
This virtual sensor is a pipeline of two stages: \texttt{FE} and \texttt{ID}.
The algorithms employed by each stage, specified by the keyword \texttt{setModel()}, are MFCC (Mel Frequency Cepstral Coefficient) and GMM (Gaussian Mixture Model), which are commonly used by voice recognition systems~\cite{lu2010jigsaw,chen2014small}.
Currently, we implement 17 data processing algorithms, including 12 for feature extraction and 5 for classification.
Although \texttt{FE} and  \texttt{ID} are the compositions of the typical pipeline, applications with more stages and parallel stages are also supported in our system, such as the EEG seizure onset detection application described in \cite{newton2009wishbone}.

Furthermore, there still a lot of complexity for green-handed developers due to they may have no idea of which sensors are strongly related to the expected output and how they are related.
To relieve this,  we propose the inference-agnostic virtual sensor.
To construct it, developers could merely provide the set of possibly related sensors and the expected output of the virtual sensor, as Fig.~\ref{fig:algorithm-agnostic-virtual-sensor} shows.
EdgeProg will first generate a simple sampling application, and developers should record the events they desired with it to obtain enough training data.
Then EdgeProg will train an inference model which reflects the relationship between the input sensors and the recorded events.
Finally, the trained model is partitioned and disseminated, similar to the other virtual sensors.

\begin{figure}
	\centering
	\begin{lstlisting}
VSensor VoiceRecog(AUTO){
	VoiceRecog.setInput(A.MIC, A.Accel_x, A.Accel_y, A.Accel_z, B.Light, B.PIR);
	VoiceRecog.setOutput(<string_t>,"open", "close");
}
	\end{lstlisting}
	\vspace{-1em}
	\caption{An example implementation code snippet of an algorithm-agnostic virtual sensor.}
	\label{fig:algorithm-agnostic-virtual-sensor}
	\vspace{-1em}
\end{figure}

\textbf{Explicit data flow.}
According to our analysis on 101 commonly-used IoT applications from several popular development websites such as DFRobot and Hackster.io, we find that about 45\% lines of code in these projects are written for data flow construction and interaction, which is a considerable proportion and increases the project complexity.
Furthermore, multi-device interaction makes the data flow more complicated due to it is conceived in the network packet construction.

In a typical IoT application, data flow starts from the production of sensor data, processed by several algorithms, then finally saved in the database or turned into a command back to the actuator IoT node.
Hence, we make the data flow explicit in these three steps.
For data production and final actuation, as illustrated in lines 2-4 of Fig.~\ref{fig:programming-language}, developers specify the data and available actions as interfaces.
For example, line 3 illustrates that three interfaces (microphone sampling, door unlocking and door opening) of a Raspberry Pi named A are used in this application.
The available interfaces of specific hardware are determined by its vendor or prototype developer.
For data processing, virtual sensors and rules directly use or call the interfaces, which results in a unified and explicit data flow.

\subsection{Code Partitioning}
\label{sec:code-partitioning}
The goal of EdgeProg's code partitioning sub-system is to divide the user input into appropriate stages and to obtain the optimal placement of each stage.
To accomplish them, we first preprocess the user input application into \textit{logic blocks}, which represent the computation stages, and generate a data flow graph of the rules to obtain a full view of the user logic as well as the stage dependency.
Afterward, as \textit{cost-aware} is one of the design goals of EdgeProg, we both formulate the latency minimizing problem and energy saving problem into a mathematical expression, then we employ an efficient solver to obtain the optimal placement of each stage.

The key insight of our partitioning algorithm is that we push the computation close to the data source as much as possible and make the best use of the computation ability of each device to achieve latency reduction.
Moreover, the optimal placement that exhibits favorable computation-transmission tradeoff could be obtained by EdgeProg benefited from the intrinsic global view of our programming language.

\subsubsection{Logic blocks and data flow graph construction}
Due to the compact nature of our programming language, there are mainly two gaps that prohibit us from further implementation and optimization.
(1) Some stages may be implicitly defined and used in the application.
For example, in Fig.~\ref{fig:programming-language}, the interface \texttt{LIGHT\_SOLAR} of device B is referenced in the rule.
Thus the stage of sensing it is necessary but being conceived from the application.
(2) The topological information is necessary for optimization, which is also implied in the application.

\begin{figure}[tbp]
	\centering
	\includegraphics[width=.8\linewidth]{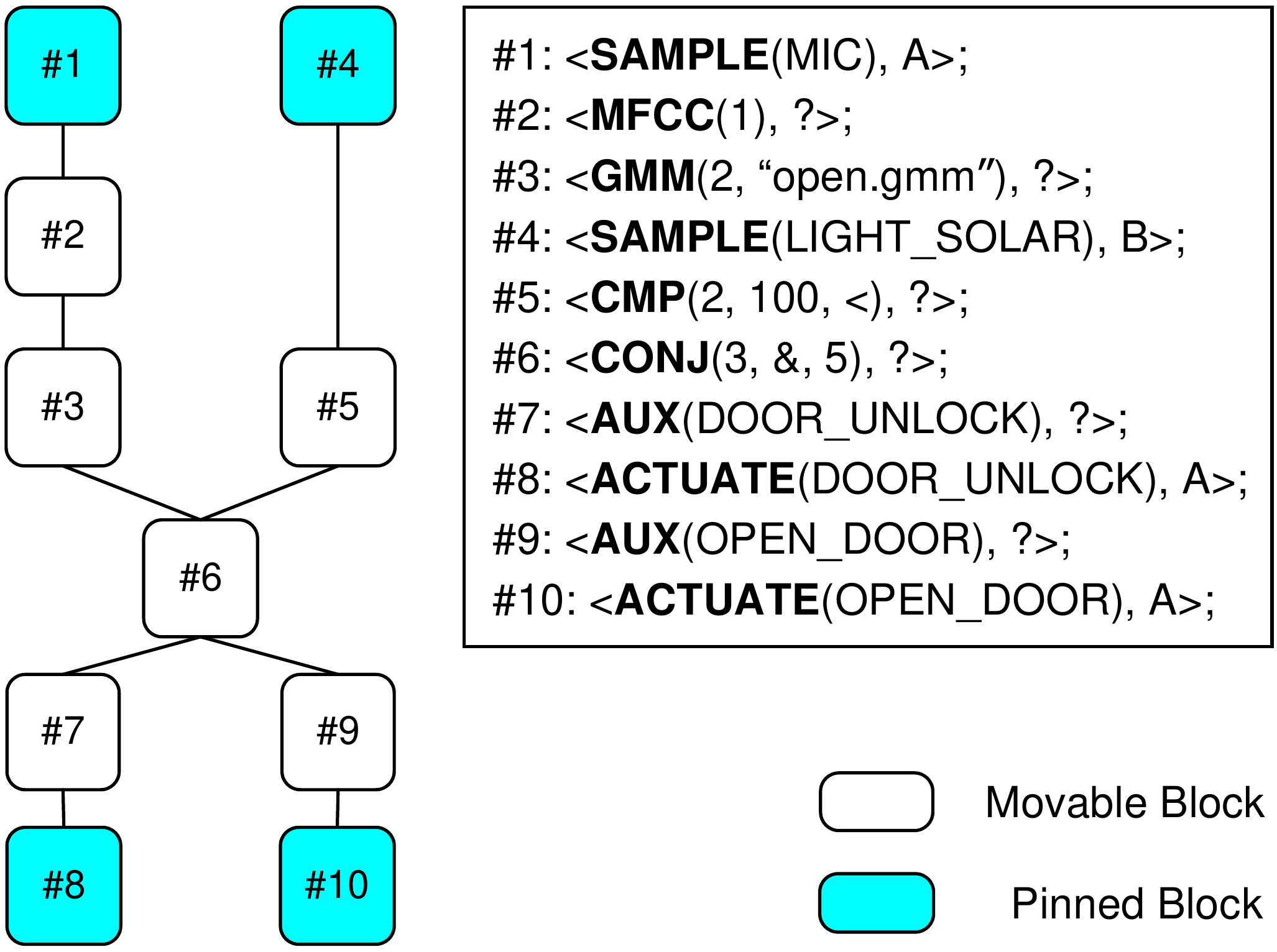}
	\vspace{-1em}
	\caption{An illustration of EdgeProg logic flow and logic block of SmartDoor application.}
	\label{fig:basic-DAG}
	\vspace{-1em}
\end{figure}

To fill up the gaps, we construct a \textit{data flow graph} of an application whose nodes are represented with \textit{logic blocks}.
A logic block is supposed to be expressive enough as an independent building block of the application, i.e., it should contain adequate information such as placement, algorithm and necessary parameters for time profiling as well as its input source for code generation.
Hence, the logic block is defined as a tuple $<$\textit{functionality}, \textit{placement}$>$, as shown in Fig.~\ref{fig:basic-DAG}.

\begin{itemize}
	\item \textit{Functionality.}
	To express the functionality, we borrow the idea of tasklet primitives from Tenet~\cite{gnawali2006tenet} such as \texttt{SAMPLE}, \texttt{ACTUATE} and \texttt{CONJ}, which provide building blocks for a wide range of data acquisition and processing tasks.
	Nevertheless, we further add the algorithms as primitives (e.g., \texttt{GMM}) to accommodate the virtual sensor deployment.
	The data source of a logic block is declared as the first argument of the primitive.
	\item \textit{Placement.}
	There are two kinds of code blocks in EdgeProg: \textit{pinned} and \textit{movable}.
	The pinned blocks are generally physical-constrained functionalities.
	For example, \texttt{SAMPLE} must be placed on the device.
	Moreover, there are also logical-constrained functionalities.
	For example, the \texttt{CONJ} is pinned to edge server to avoid unnecessary device-to-device traffic, which will lead to sub-optimal partition.
	Hence, the placement is fixed for a pinned block, and we use its corresponding device alias in the logic block.
	The placement of a movable block, which is potentially deployed on the device or edge server, is denoted with the question mark to express the uncertainty.
\end{itemize}

Generally, the logic blocks could be inferred from the EdgeProg program.
Taking the program in Fig.~\ref{fig:programming-language} as an example, each stage of the VSensor in the implementation part (i.e., \texttt{FE}, \texttt{ID}) is transformed to a logic block.
Except for these explicitly declared logic blocks, some blocks are also necessary for a complete graph but implicitly conceived in the user application.
In order to complete the data flow graph with the intrinsic blocks, we analyze all the rules defined in the \texttt{Rule} part with the following strategies:
\begin{itemize}
	\item For conditions exploiting virtual sensors in the IF statement, we refer to the \texttt{Implementation} part to obtain the staging pipeline and insert \texttt{SAMPLE} blocks for the input.
	\item For conditions that only compare sensor values, we convert it into two stages: \texttt{SAMPLE} and \texttt{CMP}.
	\item We use a \texttt{CONJ} block representing the conjunction of all the conditions in the IF statement.
	\item For each action in the THEN statement, we use two blocks: an auxiliary movable block \texttt{AUX} representing it is edge-triggered or local-triggered and a pinned block \texttt{ACTUATE} representing the action.
\end{itemize}

Then the data flow graph could be constructed as a directed acyclic graph (DAG) $G(V, E)$ whose vertices represent the logic blocks and edges represent there exist a data flow, as Fig.~\ref{fig:basic-DAG} illustrates.


\subsubsection{Formulation of optimal partitioning problem}
\label{sec:problem-formulation}
With the help of the data flow graph $G(V, E)$, we formulate the optimal partitioning problem as a numerical optimization problem.
The resulting optimal partition could be viewed as assigning each logic block to its most preferable computational device.
We would like to borrow the existing code partitioning algorithm proposed in Wishbone~\cite{newton2009wishbone} to solve our problem.

\begin{itemize}
	\item \textit{Node weights.} The weights of vertices in the graph represent the processing time or energy consumption of the corresponding logic block. In EdgeProg, the weight of a movable block is two-fold: the local and edge-server processing time/energy. While in Wishbone, each vertex in the graph only has one weight.
	\item \textit{Optimization goal.} The optimization goal of Wishbone is minimizing the sum of computational budgets and network bandwidth. Nevertheless, EdgeProg focuses on the cost of executing the application (i.e., latency or energy), which makes the Wishbone formulation no longer suitable for our problem.
\end{itemize}


In EdgeProg, we support two optimization goals: execution time or energy consumption of an edge-device integrated application, and users could choose one of the goals on their demand.
We next present the formulation of latency and energy optimization, respectively.

\textbf{Optimizing execution time.}
Minimizing the task execution latency leads to minimizing the length of the longest path in the data flow graph.
We define a full path of the data flow graph $G(V, E)$ as the path from a source vertex to a sink vertex, denoted as $\pi$.
We use $len(\pi)$, $\delta(\pi)$ and $\Pi(G)$ to represent the length of path $\pi$, the number of vertices in path $\pi$, and the set of all full paths in graph $G$.
Thus, our optimization goal is thus denoted as:
\begin{equation}
	\min \max_{\pi \in \Pi(G)} len(\pi).
\end{equation}
In order to further demonstrate $len(\pi)$, we first introduce a binary indicator $X_{b_is}$ to demonstrate the placement as:
\begin{equation}
	X_{b_is} =
	\begin{cases}
		1 & \text{logic block } b_i \text{ is assigned to device } s\\
		0 & \text{logic block } b_i \text{ is not assigned to device } s
	\end{cases},
\end{equation}
where $s \in S_i$, and $S_i$ denotes the set of all possible devices that could place the $i$-th logic block (i.e., $b_i$).
Thus, the sum of computing and transmitting latency across all possible placements of a full path, $len(\pi)$, could be expressed as:
\begin{equation}
	\label{equ:obj-origin}
	len(\pi) = \sum_{i=1}^{\delta(\pi)}\sum_{s \in S_i}X_{b_is}T_{b_is}^{C} + \sum_{i=1}^{\delta(\pi)-1}\!\!\!\!\!\!\sum_{\mbox{\tiny$\begin{array}{c}
				s\!\in\!S_i\\s^{\prime}\!\in\!S_{i^{\prime}}
			\end{array}$}}\!\!\!\!\!\!X_{b_is}X_{b_{i^{\prime}}s^{\prime}}T^{N}_{b_i s s^{\prime}},
\end{equation}
where $i$, $i^\prime$ are the adjacent vertices in path $p$ (i.e., $i^\prime$=$i$+1).
We use $T_{b_is}^{C}$ to denote the data processing cost of the $i$-th block on device $s$, and $T^{N}_{b_i s s^{\prime}}$ to represent the data transmission time between block $b_i$ of device $s$ and block $b_{i^{\prime}}$ of device $s^{\prime}$.
We assume that the data transmission time is negligible if the two consecutive logic blocks are placed on the same device.
Thus we have:
\begin{equation}
	\label{equ:network}
	T^{N}_{b_i s s^{\prime}} =
	\begin{cases}
		\left \lceil\frac{q_{ii^{\prime}}}{r_{ii^{\prime} k}} \right \rceil t_{ii^{\prime} k}& s \neq s^{\prime}\\
		0 & s = s^{\prime}
	\end{cases},
\end{equation}
where $q_{ii^{\prime}}$ denotes the data size being transmitted on edge ($i$, $i^{\prime}$).
$r_{ii^{\prime} k}$ is a protocol-specific metric representing the maximum packet payload of protocol $k$, e.g., the $r_{ii^{\prime} k}$ of 6LowPAN network is 122 bytes.
Furthermore, the per-packet transmission time is given by $t_{ii^{\prime} k}$, which is profiled and predicted by our network profiler detailed in Section~\ref{sec:architecture}.

\textbf{Optimizing energy consumption.}
Different from the latency formulation, optimizing energy consumption should consider the costs of all the edges and vertices rather than consider the longest path only.
Thus, the optimization goal is formulated as:
\begin{equation}
	\label{equ:obj-origin-energy}
	\mathop{\arg\min}_X \sum_{i=1}^{|V|}\sum_{s \in S_i}X_{b_is}E_{b_is}^{C} + \sum_{i=1}^{|V|}\!\!\!\!\sum_{\mbox{\tiny$\begin{array}{c}
				s\in S_i\\s^{\prime}\!\in\!S_{i^{\prime}}
			\end{array}$}}\!\!\!\!\!\!X_{b_is}X_{b_{i^{\prime}}s^{\prime}}E^{N}_{b_i s s^{\prime}},
\end{equation}
where $|V|$ stands for the number of vertices in $G(V, E)$.
The data processing energy $E_{b_is}^{C}$ and transmission energy $E^{N}_{b_i s s^{\prime}}$ could be derived by:
\begin{equation}
	\label{equ:energy}
	\begin{cases}
		E_{b_is}^{C} = T_{b_is}^{C} P^C_{s}\\
		E^{N}_{b_i s s^{\prime}} = T^{N}_{b_i s s^{\prime}} (P^{TX}_{s} + P^{RX}_{s^\prime})
	\end{cases},
\end{equation}
where $P^C_{s}$ stands for the average power (in mW) of device $s$ for computation.
$P^{TX}_{s}$ and $P^{RX}_{s^\prime}$ represent the average power for TX operation of device $s$ and RX operation of device $s^\prime$, respectively.
It is worth noting that, we only consider the energy consumption of IoT devices.
The energy consumed by edge devices are ignored (i.e., $P^C_{s}$, $P^{TX}_{s}$ and $P^{RX}_{s^\prime}$ are set to 0) due to the edge devices are mostly AC-powered.

\subsubsection{Solution of optimal partitioning problem}
The objective formulations of EdgeProg optimization problem (i.e., Equ.~(\ref{equ:obj-origin}) and (\ref{equ:obj-origin-energy})) are \textit{quadratic programming} (QP) problems, which are shown to be NP-hard~\cite{eidenbenz2016task}.
The state-of-the-arts employ heuristic algorithms to solve it efficiently.
For example, the most recent work~\cite{khare2019linearize} utilizes a breadth-first greedy search algorithm to solve it.
While we prefer a solving method, which is less prone to local optima and the method's scalability against problem size is also a necessary property.

Inspired by McCormick Envelopes relaxation~\cite{mitsos2009mccormick}, we re-formulate the objectives and constraints of EdgeProg to conform to the formulation of integer linear programming problem (ILP), which could be efficiently solved by the standard solver, e.g., \texttt{lp\_solve}.
\iftechreport
We compare the solving time of the QP and ILP formulations in Appendix~\ref{sec:appendix-gurobi}.
\else
We compare the solving time of the QP and ILP formulations in our technical report~\cite{li2021edgeprogtechreport}.
\fi
Results show that the ILP formulation is more scalable than QP in terms of solving time.

Take the latency optimization problem (Equ.~(\ref{equ:obj-origin})) of EdgeProg as an example.
We first convert the quadratic objective function to linear one by introducing an auxiliary variable $\epsilon_{is s^{\prime}}$=$X_{b_{i}s}$$\cdot$ $X_{b_i{^{\prime}}s^{\prime}}$ to replace the quadratic term $X_{b_{i}s}$$\cdot$ $X_{b_i{^{\prime}}s^{\prime}}$ in Equ.~(\ref{equ:obj-origin}).
Moreover, the presence of $\epsilon_{is s^{\prime}}$ causes the introduction of these constraints:
\begin{align}
	&(\forall i\!\in \delta(p)\!-\!1,s\!\in S_i, s^{\prime}\!\in\!S_{i^{\prime}})\ \ \epsilon_{is s^{\prime}} \geq 0, \label{equ:constr-linear1}\\
	&(\forall i\!\in \delta(p)\!-\!1,s\!\in S_i, s^{\prime}\!\in\!S_{i^{\prime}})\ \ \epsilon_{is s^{\prime}} \leq X_{b_is}, \label{equ:constr-linear2} \\
	&(\forall i\!\in \delta(p)\!-\!1,s\!\in S_i, s^{\prime}\!\in\!S_{i^{\prime}})\ \ \epsilon_{is s^{\prime}} \leq X_{b_{i^{\prime}}s^{\prime}}, \label{equ:constr-linear3}\\
	&(\forall i\!\in \delta(p)\!-\!1,s\!\in S_i, s^{\prime}\!\in\!S_{i^{\prime}})\ \ \epsilon_{is s^{\prime}}\! +\! 1 \!\geq\! X_{b_{i}s}\!+X_{b_{i^{\prime}}s^{\prime}}. \label{equ:constr-linear4}
\end{align}
It can be observed that all the above four constraints are linear.
Whereas our objective function is still in a minimax shape, which needs further transformation.
We thus introduce another auxiliary variable $z$ and convert the inner \texttt{max} function to a set of constraints to make it follow standard ILP formulation.
The rewritten ILP objective function is:
\begin{alignat}{3}
	&\text{\textbf{Objective:}} \quad&&\mathop{\arg\min}_X  \quad z \label{equ:obj-final}\\
	&\text{\textbf{Subject to:}} \quad&& \nonumber \\ 
	&z \geq \sum_{i=1}^{\delta(\pi)}\sum_{s \in S_i}X_{b_is}T_{b_is}^{C} &+& \sum_{i=1}^{\delta(\pi)-1}\!\!\!\!\!\!\sum_{\mbox{\tiny$\begin{array}{c}
				s\!\in\!S_i\\s^{\prime}\!\in\!S_{i^{\prime}}
			\end{array}$}}\!\!\!\!\!\!\epsilon_{is s^{\prime}}T^{N}_{b_i s s^{\prime}}, \forall \pi \in G. \label{equ:constr-minimax}
\end{alignat}

Furthermore, we add constraints for $X_{b_is}$ to ensure each logic block is appointed to a specific device.
\begin{align}
	\label{equ:constr-appoint}
	\sum_{s \in S_i}X_{b_is}=1, \forall i \in G.
\end{align}
Thus, any optimal solution of Equ.~(\ref{equ:obj-final}) subject to (\ref{equ:constr-linear1})-(\ref{equ:constr-linear4}), (\ref{equ:constr-minimax}) and (\ref{equ:constr-appoint}) will be the optimal partition of the input application.

Similarly, the objective of energy optimization problem (Equ.~\ref{equ:obj-origin-energy}) is transformed as:
\begin{equation}
	\label{equ:obj-energy-final}
	\mathop{\arg\min}_X \sum_{i=1}^{|V|}\sum_{s \in S_i}X_{b_is}E_{b_is}^{C} + \sum_{i=1}^{|V|}\!\!\!\!\sum_{\mbox{\tiny$\begin{array}{c}
				s\!\in\!S_i\\s^{\prime}\!\in\!\!S_{i^{\prime}}
			\end{array}$}}\!\!\!\epsilon_{i s s^\prime} E^{N}_{b_i s s^{\prime}},
\end{equation}
along with the formulations of $E_{b_is}^{C}$ and $E^{N}_{b_i s s^{\prime}}$ (Equ.~\ref{equ:energy}), and the constraints similar to Equ.~(\ref{equ:constr-linear1})-(\ref{equ:constr-linear4}) and (\ref{equ:constr-appoint}).

\subsection{Executable Generator}
\label{sec:executable-generator}

\begin{figure}
	\centering
	\begin{lstlisting}
PROCESS_THREAD(cond1_process, ev, data){
	static struct etimer et_cond1;
	PROCESS_BEGIN();
	etimer_set(&et_cond1, COND1_INTERVAL);
	while(1){
		PROCESS_YIELD();
		if (etimer_expired(&et_cond1)){
			(*@ \textbf{(do the jobs of logic blocks)} @*)
			process_post(&send_process, send_evt, &to_send1);(*@\label{loc:code-generation-send-process}@*)
			etimer_reset(&et_cond1);
		}
	}
	PROCESS_END();
}
	\end{lstlisting}
	\vspace{-1em}
	\caption{Code snippets of the functioning process of EdgeProg.}
	\vspace{-1em}
	\label{fig:code-generation}
\end{figure}

The executable generation process in EdgeProg contains two steps: 
(1) constructing pieces of compilable code from the optimal partition and the logic blocks,
and (2) compiling the code to platform-specific executables.

Benefited by the cross-platform nature of Contiki OS
, we could generate the code for the edge server (mostly Linux-compatible hardware) as well as sensing devices in a similar manner.
Then EdgeProg compiles them using the platform-specific toolchains provided by Contiki based on \texttt{msp430-gcc} for TelosB and \texttt{gcc-linaro-arm} for Raspberry Pi.
The only difference our generator should take care of is the different libraries included and sampling APIs used for distinct platforms.
Hence, we focus on how to generate compilable code that runs efficiently.

As we mentioned in the last section, the logic blocks are designed to be expressive enough to act as a building block of an application, and hence they are transformed to a function into the final compilable code.
The most difficult issue is how to organize the function calls in the generated code.
The intuitive approach to accommodate the event-driven kernel and the protothread technique of Contiki OS is to arrange all the logic blocks assigned to the same placement in a protothread and send/receive data if the next block is assigned to another device.
This simple design raises performance drawbacks.
The generated protothread could be too long with this design, which degrades the system performance due to the non-preemptive scheduling of Contiki\footnote{Contiki supports preemptive multi-threading as an optional library, while it requires additional multiple stack allocation which is stressful for low-end devices such as TelosB. Hence we do not adopt this scheme.}.
Generating one protothread of one block is also not efficient because short protothread incurs much process switching overhead, which will also harm the performance.

Our approach is based on a code template of Contiki necessaries and a send thread with receive callback.
The functioning protothreads are generated from graph fragments of the optimized DAG, and the code snippet is illustrated in Fig.~\ref{fig:code-generation}.
The fragments of each device are obtained by leveraging a depth-first traverse of the logic blocks of the DAG which ends at the placement-changing point.
Then we assemble a protothread with one fragment by calling functions of the logic blocks.
At the end of a thread, it issues an event to the send thread (e.g., line~\ref{loc:code-generation-send-process} of Fig.~\ref{fig:code-generation}) for data transmission and yields for other threads.
Moreover, based on our time profiling, the graph fragments could be further segmented if it contains several time-consuming tasks for system health.

\section{Evaluation}
\label{sec:evaluation}
In this section, we evaluate the performance of EdgeProg in various aspects.

\subsection{Experiment Setup}
\label{sec:evaluation-experiment-setup}

\begin{table*}
	\centering
	\caption{Implemented benchmark applications.}
	\iftechreport\else
	\vspace{-1em}
	\fi
	\label{tab:evaluation-overall}
	\begin{tabular}{|c|c|c|c|c|}
		\hline
		\textbf{Name}  &\textbf{Application}			&	\textbf{Sensor}	&	\textbf{\# Operators} & \textbf{Algorithms}\\ \hline
		\textbf{Sense} &Outlier Detector~\cite{kumar2007harbor,marcelloni2009efficient} &Temp., Light& 8 & Average, Matrix multiplication, LEC compression\\ \hline
		\textbf{MNSVG} &Weather Forecasting~\cite{heo2017rt}&Temp., Humidity & 4 &MNSVG \\ \hline
		\textbf{EEG}   &Seizure Onset Detect.~\cite{shoeb2006detecting} & EEG & 80 &Wavelet decomposition, SVM \\ \hline
		\textbf{SHOW}  &Smart Handwriting~\cite{lin2018show} & Accel. & 13 & FFT, Random forest \\ \hline
		\textbf{Voice} &Speaker Count~\cite{xu2013crowd++} & MIC & 10 & MFCC, Pitch estimation, Unsupervised clustering \\
		\hline
	\end{tabular}
	\vspace{-1em}
\end{table*}

First, we introduce the benchmarks we used and baselines we compared in our evaluation.

\textbf{Benchmarks.}
We summarize the five macro-benchmarks to evaluate our EdgeProg in Table \ref{tab:evaluation-overall}: two sensing applications and three real-world applications.
The \#operators column in Table~\ref{tab:evaluation-overall} indicates the number of operational logic blocks of each benchmark.

\begin{itemize}
	\item \textit{Sense.} A common sensing application with outlier detection using algorithms proposed in~\cite{kumar2007harbor} and data compression using the LEC algorithm~\cite{marcelloni2009efficient}.
	\item \textit{MNSVG.} A weather forecast application using an MNSVG model proposed in~\cite{heo2017rt} to predict temperature and humidity values.
	\item \textit{EEG.} Using the EEG signal to detect seizures~\cite{shoeb2006detecting} , taken from Wishbone~\cite{newton2009wishbone}. 
	It employs ten parallel channels to process the EEG signal with seven order wavelet decomposition in each channel.
	\item \textit{SHOW.} Detecting and classifying the trajectory of the device with IMU information and random forest algorithm~\cite{lin2018show}.
	\item \textit{Voice.} Counting the number of speakers with signal processing and clustering algorithms~\cite{xu2013crowd++}.
\end{itemize}

\textbf{Baselines Definition.}
Here we describe the state-of-the-art edge(cloud)-device interactive system alternatives that we use to illustrate the advantages of EdgeProg.
\begin{itemize}
	\item \textit{RT-IFTTT}~\cite{heo2017rt}. The server does all of the computation. IoT devices only need to report the sensor value or take actions under the server's command.
	\item \textit{Wishbone(0.5, 0.5)}~\cite{newton2009wishbone}. Wishbone is a partitioning system for sensornet applications whose goal is to minimize a combined objective of CPU and network workload, which could be formulated with two weights as ($\alpha CPU + \beta Net$). 
	Here (0.5, 0.5) stands for $\alpha=\beta=0.5$, which indicates CPU and network are of equal importance in this baseline. 
	\item \textit{Wishbone(opt.)}. During our preliminary experiment, we notice that better latency performance could be achieved by altering the $\alpha$ and $\beta$ parameters.
	Hence, we conduct evaluations by tuning the parameters with 0.1 step, and record the best performance as this baseline.
\end{itemize}

\subsection{Latency Reduction}
\label{sec:evaluation-latency}
\begin{figure*}[htbp]
	\centering
	\subfigure[Execution time under Zigbee network.]{
		\begin{minipage}[t]{0.5\linewidth}
			\centering
			\includegraphics[width=.9\linewidth]{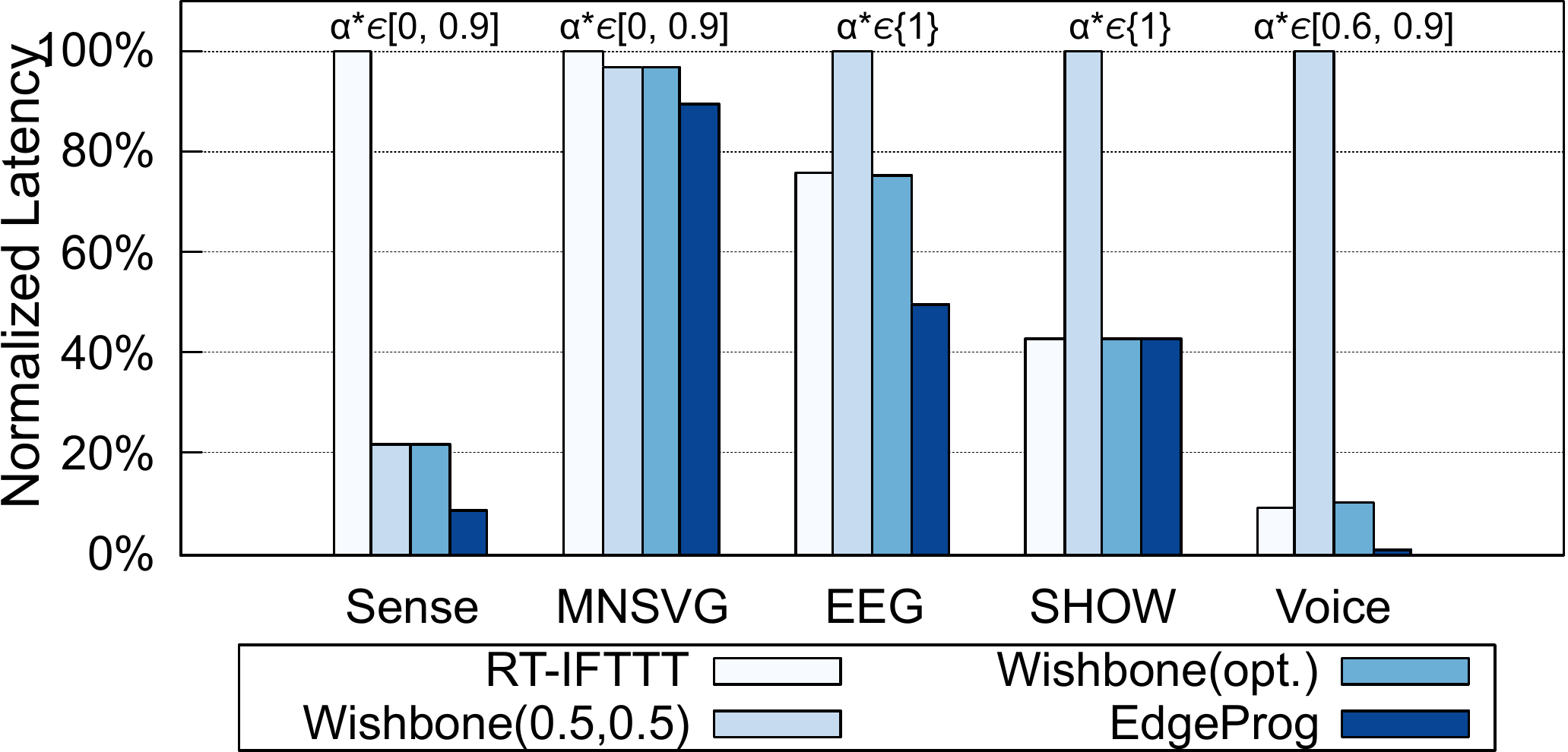}
		\end{minipage}%
	}%
	\subfigure[Execution time under WiFi network.]{
		\begin{minipage}[t]{0.5\linewidth}
			\centering
			\includegraphics[width=.9\linewidth]{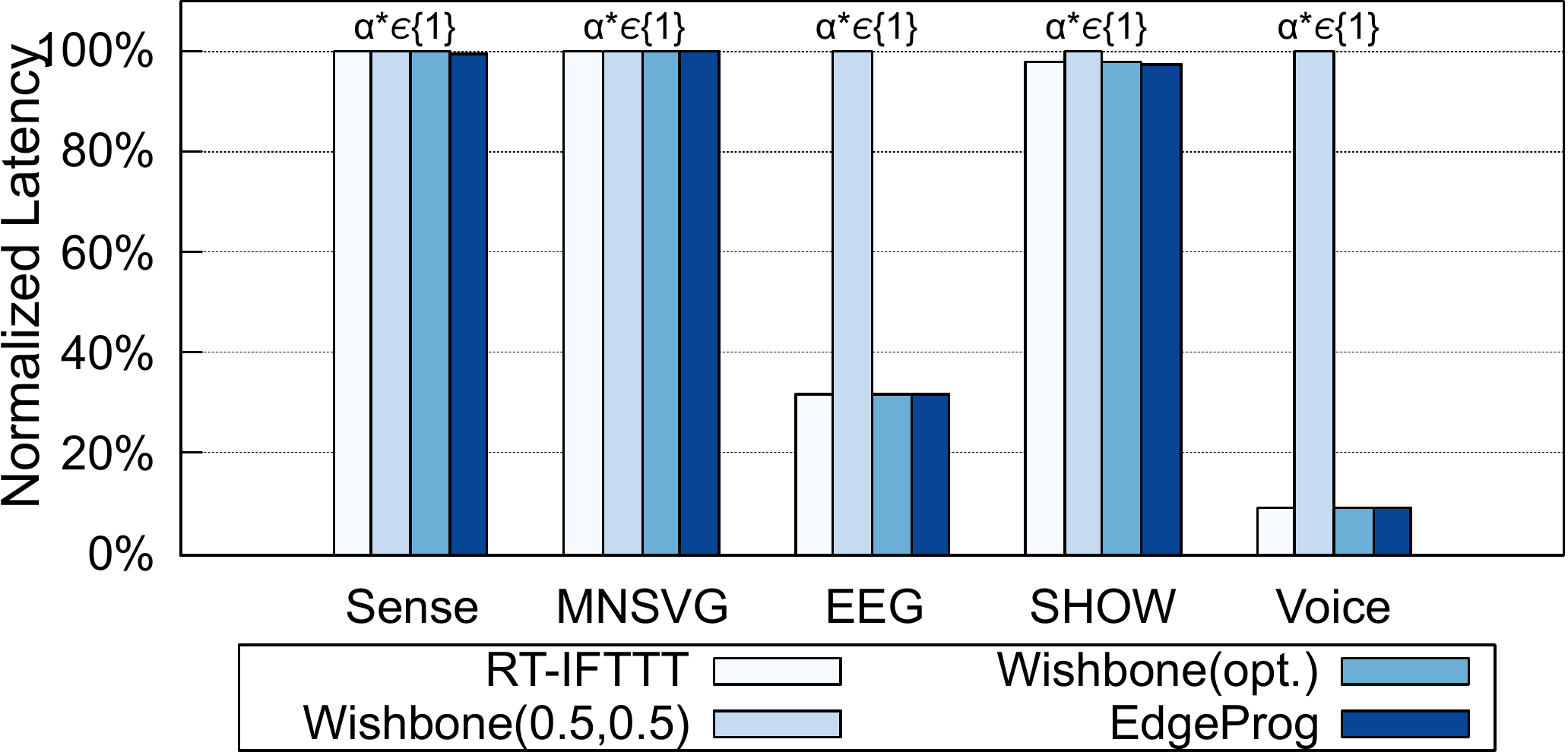}
		\end{minipage}%
	}%
	\vspace{-1em}
	\caption{Latency measurements normalized to the worst-performed baseline. EdgeProg reduces the task latency by 18.2\% compared with Wishbone(opt.) and 31.0\% with RT-IFTTT on average. The optimal range of $\alpha$ of Wishbone(opt.) for each benchmark is labeled on the top.}
	\label{fig:overall-results-latency}
	\vspace{-1em}
\end{figure*}

Fig.~\ref{fig:overall-results-latency} depicts the task makespan of five macro-benchmarks under Zigbee (on TelosB node) and WiFi (on Raspberry Pi) network.
We use a laptop with a 2.8GHz i7-7700HQ CPU and 16GB memory as our edge server.
EdgeProg achieves a 20.96\% reduction on average across all settings, and up to 99.05\% reduction in Voice benchmark compared with Wishbone(0.5, 0.5).
Moreover, we have two main observations according to the results:

\begin{figure*}[htbp]
	\centering
	\subfigure[Sense]{
		\begin{minipage}[t]{0.3\linewidth}
			\centering
			\includegraphics[width=.9\linewidth]{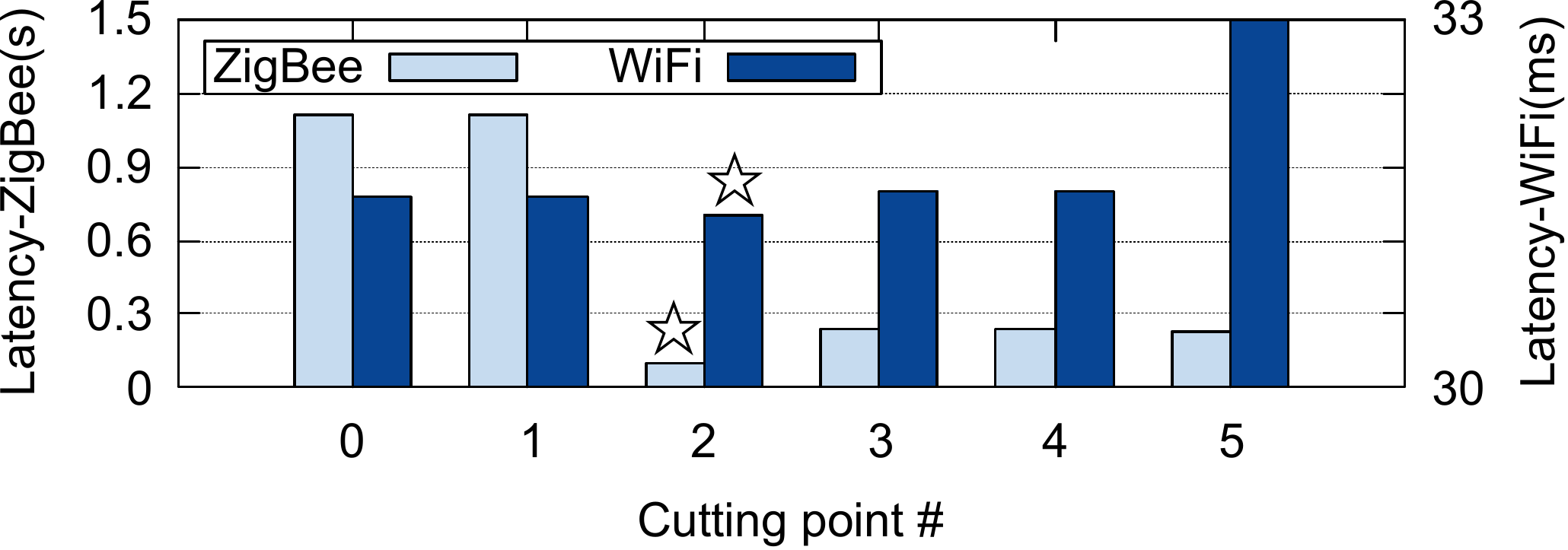}
		\end{minipage}%
	}%
	\subfigure[MNSVG]{
		\begin{minipage}[t]{0.3\linewidth}
			\centering
			\includegraphics[width=.9\linewidth]{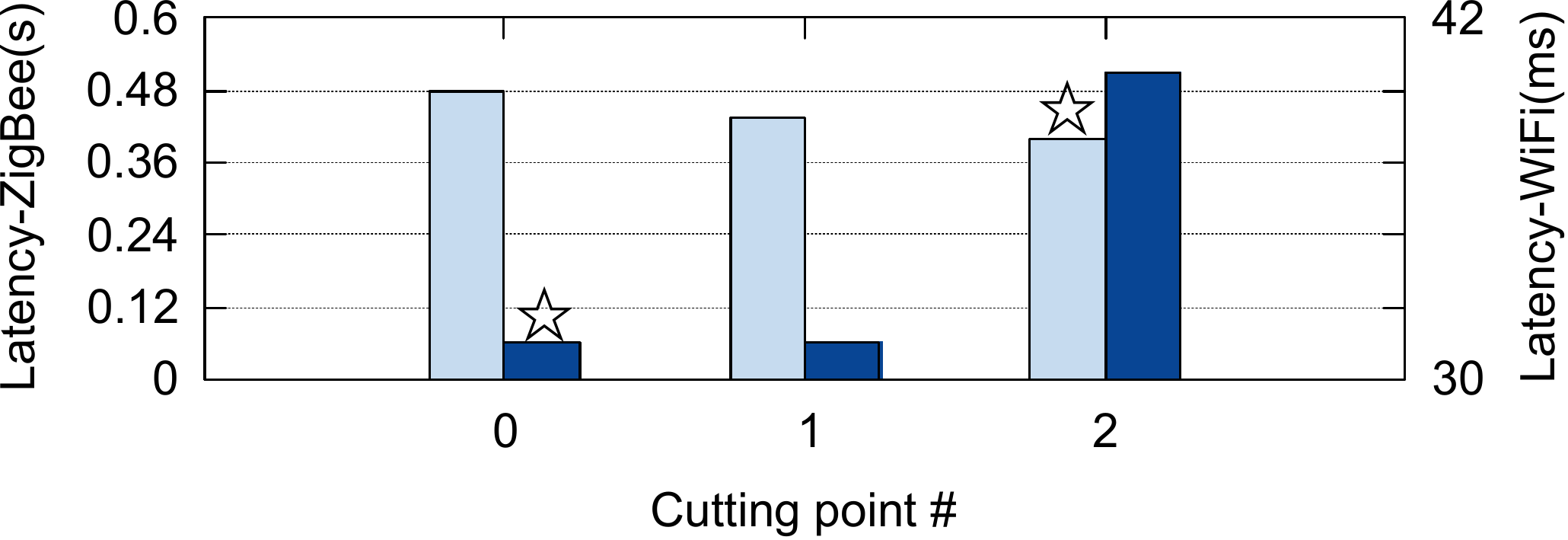}
		\end{minipage}%
	}%
	\subfigure[EEG]{
		\begin{minipage}[t]{0.3\linewidth}
			\centering
			\includegraphics[width=.9\linewidth]{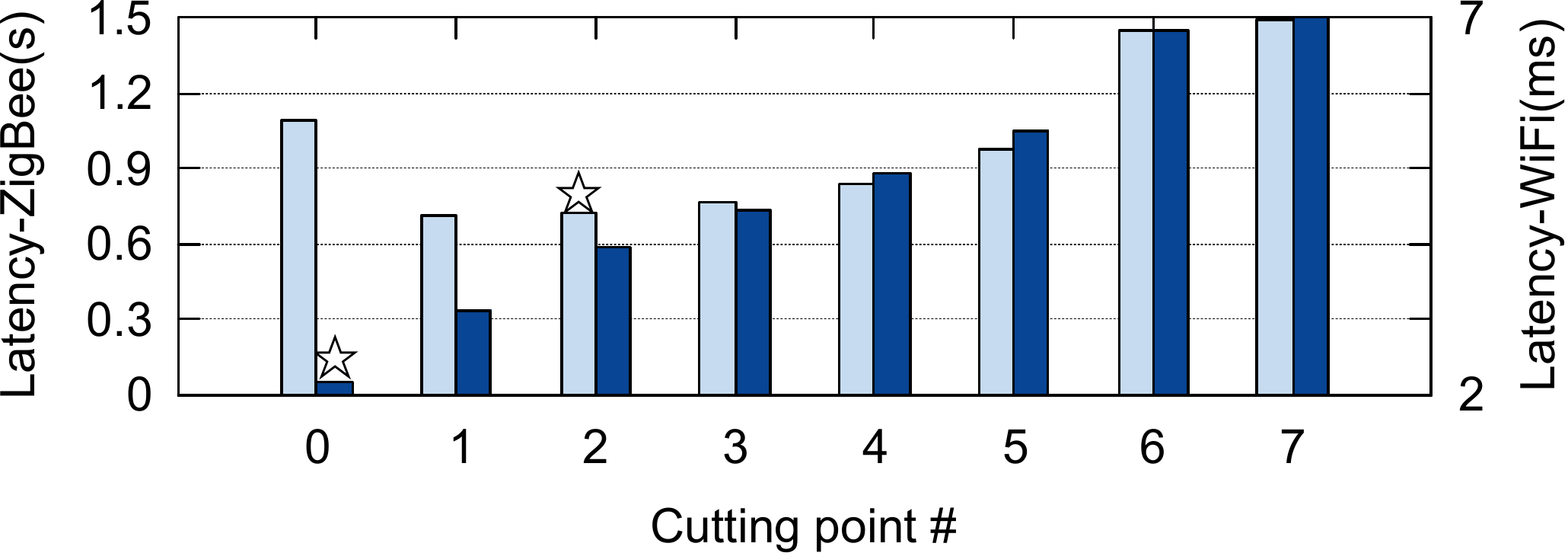}
		\end{minipage}%
	}%
	
	\vspace{-1em}
	
	\subfigure[SHOW]{
		\begin{minipage}[t]{0.3\linewidth}
			\centering
			\includegraphics[width=.9\linewidth]{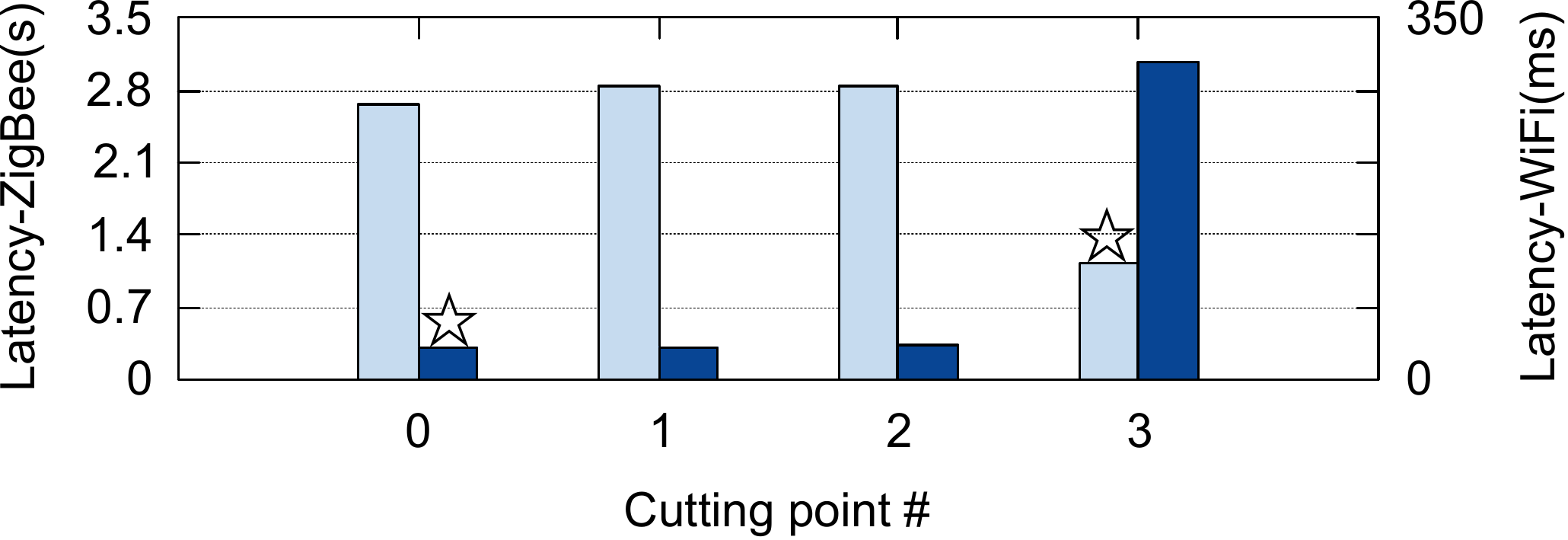}
		\end{minipage}%
	}%
	\subfigure[Voice]{
		\begin{minipage}[t]{0.3\linewidth}
			\centering
			\includegraphics[width=.9\linewidth]{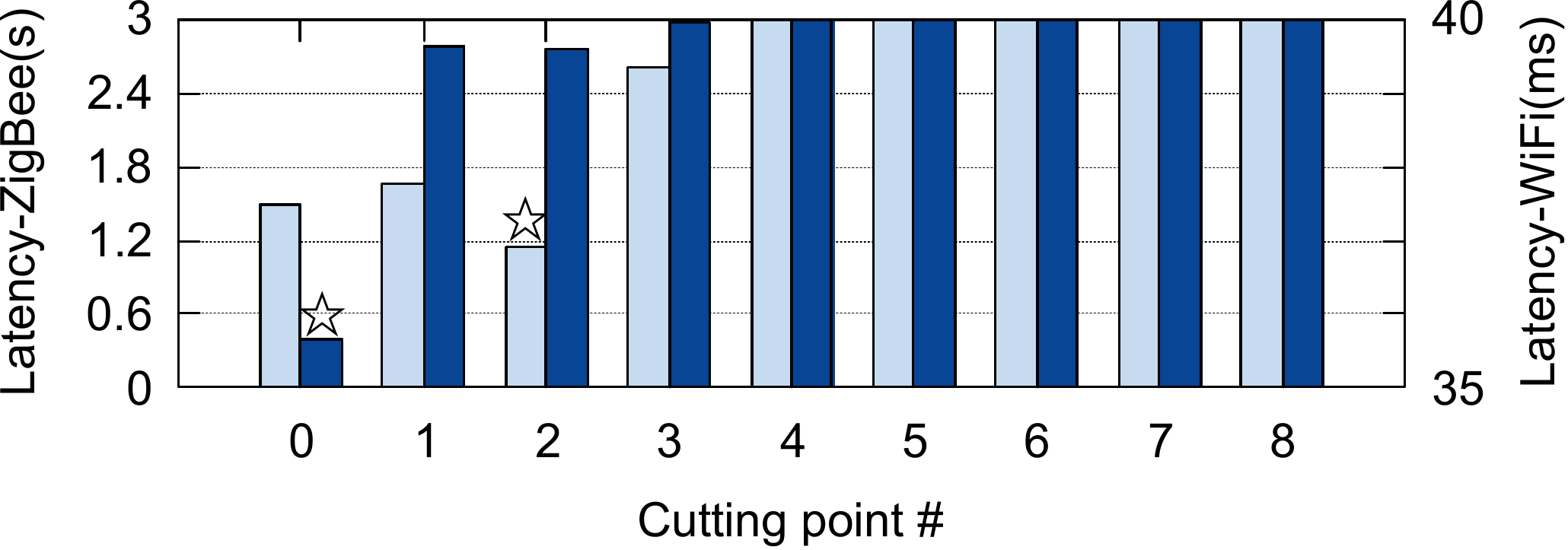}
		\end{minipage}%
	}%
	\centering
	\vspace{-1em}
	\caption{Latency of each macro-benchmark at available cutting points under both networks. Cutting points are arranged to assure that fewer operators are executed locally when point number gets bigger. We omit the bigger cutting points part of Voice and EEG due to the continuous growth of latency.}
	\vspace{-0.5em}
	\label{fig:cut-points}
\end{figure*}

(1) Speed up percentage varies considerably among benchmarks.
For example, EdgeProg surpasses the baselines for Voice and EEG benchmarks under both settings while falls flat for MNSVG.
This variation mainly due to the computation complexity and network demands of each benchmark.
As illustrated in Table \ref{tab:evaluation-overall}, EEG is the most complex one with 80 operators, which promises a larger optimization space to reduce the latency.
Furthermore, each order of its wavelet decomposition halves input data, which reduces the transmission time of its output and makes it more profitable to local execution.
Nevertheless, EdgeProg struggles against SHOW with 13 operators under WiFi, mainly due to the parallel layout of its operators, which leads to fewer valid cut points to partition.
As for MNSVG, a small number of its operators results in its available cut points is only three.
Under this circumstance, EdgeProg still captures the best cut point for ZigBee, which is neglected by baseline methods.
In summary, data-reduction algorithms contribute more to latency reduction.

(2) EdgeProg under ZigBee network outperforms under WiFi.
Under the ZigBee network, EdgeProg reduces the makespan by 30.96\%, 45.80\% and 18.19\% compared with three baselines, individually.
Nevertheless, reduction percentages drop to 0.07\%, 30.58\% and 0.13\% when using WiFi.
This is because 
 the WiFi network is much faster than Zigbee, which leads to a short networking time, and the data processing time/energy becomes the dominant fraction in the algorithm. 
Furthermore, the IoT device we used for WiFi (Raspberry Pi) has better computing power than the device we used for Zigbee (TelosB), which leads to a smaller difference in the data processing performance between the two partitions.
Hence, among the benchmarks, both the networking time and the computing time become closer to the sub-optimal ones under WiFi. 
Therefore, the optimization space of EdgeProg becomes smaller under WiFi network, which finally causes a smaller performance gain under WiFi than Zigbee. 
It is worth noting that although the performance of each approach is close to each other under WiFi, EdgeProg always obtains the optimal partition for each benchmark.

To further study the above observation, we established a ground truth by exhaustively running each benchmark at every available cutting points on our testbed. 
Fig.~\ref{fig:cut-points} illustrates the results.
The star icons indicate EdgeProg's choice for the best cutting points. 
We can infer from the figures that as the network speed grows, data transmission time decreases and data processing time becomes dominant.
Hence, optimization algorithms prefer to offload tasks at early stages, which could be deduced from that the star icons on WiFi bars are more to the left than ZigBee ones.
Consequently, the dominant strategies are more concentrated on the left, which means the decrease of optimization space and leads to closer performance among baselines.

\subsection{Energy Saving}
\begin{figure*}[htbp]
	\centering
	\subfigure[Energy consumption under Zigbee network.]{
		\begin{minipage}[t]{0.5\linewidth}
			\centering
			\includegraphics[width=.9\linewidth]{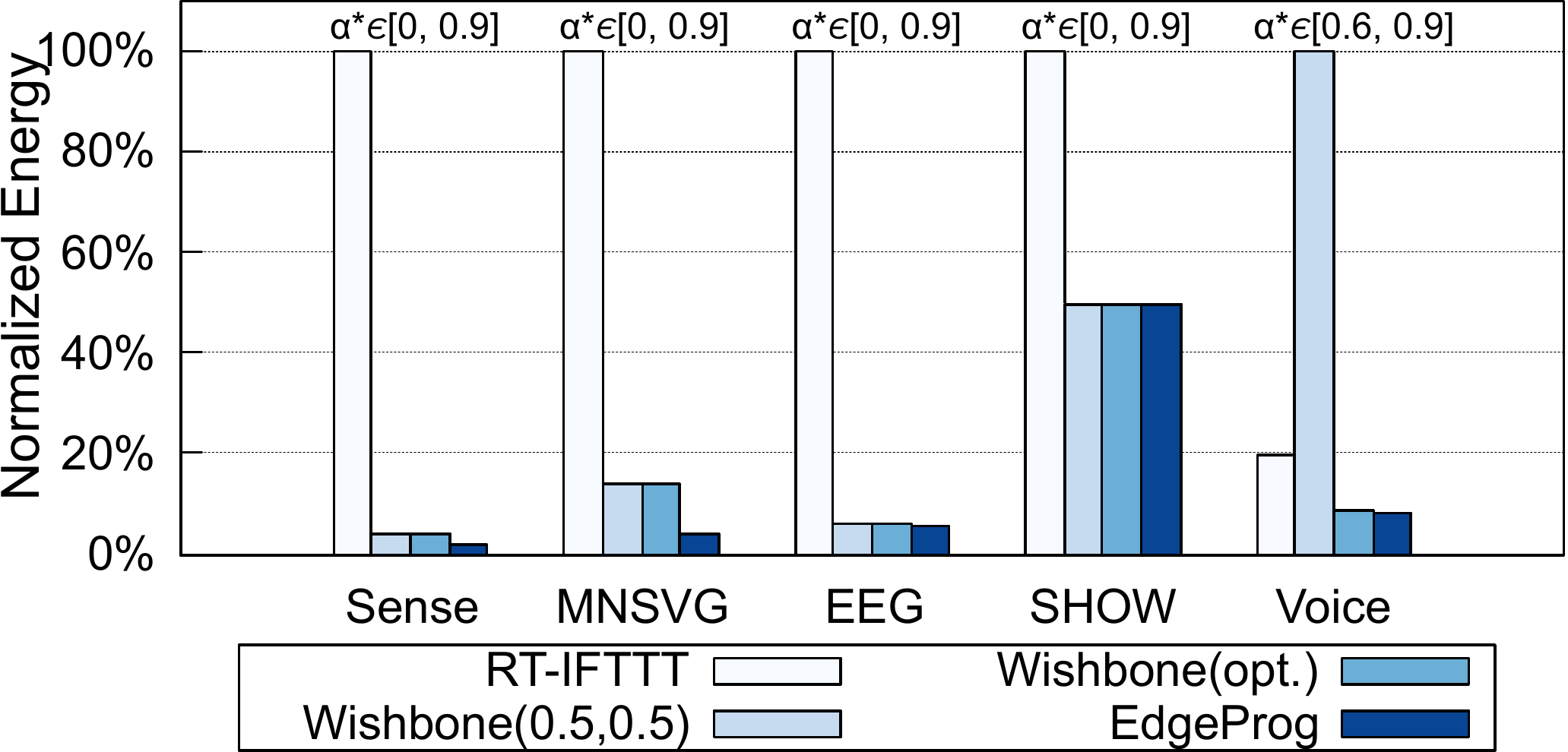}
		\end{minipage}%
	}%
	\subfigure[Energy consumption under WiFi network.]{
		\begin{minipage}[t]{0.5\linewidth}
			\centering
			\includegraphics[width=.9\linewidth]{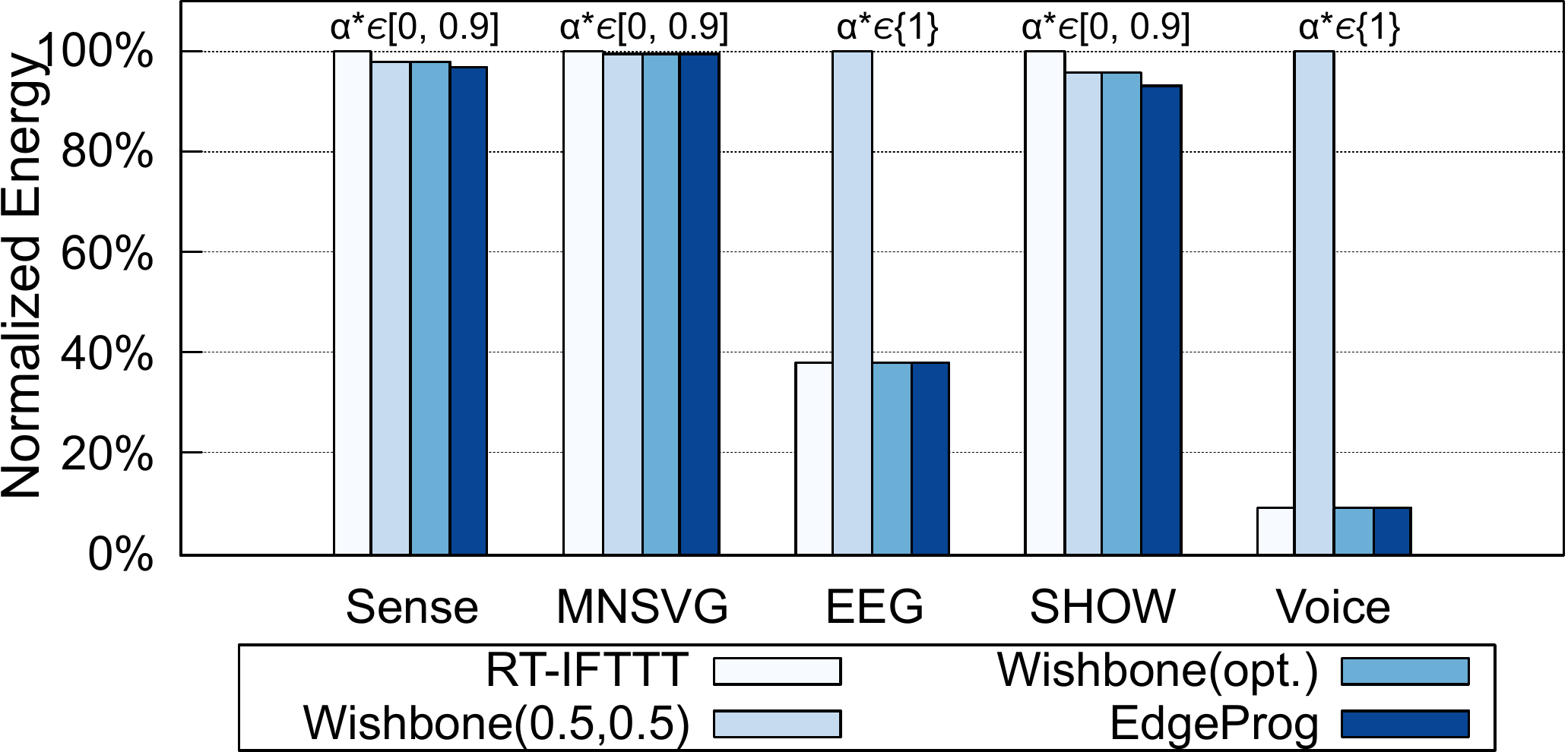}
		\end{minipage}%
	}%
	\vspace{-1em}
	\caption{Energy consumption comparison results (normalized to the worst-performed baseline). Compared with Wishbone(opt.) and RT-IFTTT, EdgeProg saves the energy by 14.8\% and 40.8\% on average. The optimal range of $\alpha$ of Wishbone(opt.) for each benchmark is labeled on the top.}
	\label{fig:overall-results-energy}
	\vspace{-1em}
\end{figure*}

Besides latency, EdgeProg could also optimize the energy consumption. 
The experiment setup is similar to the ones in Section~5.2, and we use the Monsoon Power Monitor to measure the energy consumption. 
As we specified in Section~\ref{sec:problem-formulation}, we do not consider the power consumed by edge devices because they are mostly AC-powered.
Fig.~\ref{fig:overall-results-energy} illustrates the evaluation results under Zigbee and WiFi networks.
EdgeProg achieves 31.48\% overall energy saving on average across all settings, and up to 98.38\% reduction in Sense benchmark compared with RT-IFTTT under Zigbee network.
We also observed that EdgeProg performs better under Zigbee network (51.60\% average reduction) than WiFi (11.37\%), and the reason is similar to the latency in Section~\ref{sec:evaluation-latency}.

Nevertheless, there are also some situations that EdgeProg seems to make no optimization to the latency or energy (e.g., the SHOW benchmark in Fig.~\ref{fig:overall-results-latency}(a)). 
The reason of EdgeProg achieves no optimization in certain benchmarks is that the comparing baselines have already achieved the optimal performance (i.e., these methods partition the application in the optimal way) in those situations. 
However, EdgeProg shows its generality for achieving optimal solutions under different setups (i.e., different optimization goals, application structures, etc.). 
For example, although the baselines reach the optimal partition for the SHOW benchmark under Zigbee network, only EdgeProg achieves the optimal solution for SHOW benchmark under WiFi network, as shown in Fig.~\ref{fig:overall-results-latency}(b) and \ref{fig:overall-results-energy}(b).

Furthermore, according to the results, we found that the optimal range of $\alpha$ (i.e., $\alpha^*$) for Wishbone(opt.) varies among benchmarks and optimization goals (latency, energy).
Wishbone~\cite{newton2009wishbone} claims that its objective $\alpha CPU+\beta Net$ could be a proxy for meaningful objectives such as energy, but the per-benchmark variation of $\alpha^*$ makes it difficult to take advantage of Wishbone in practice.
More specifically, besides the optimization goal, the $\alpha^*$ is influenced by the task type and device characteristic.
1) \textit{Influence of tasks type}.
For example, in Fig.~\ref{fig:overall-results-latency}(a), the $\alpha_{Sense}^*$ ($\alpha^*$ of Sense benchmark) tend to be small while the $\alpha_{EEG}^*$ is 1 when minimizing latency.
EEG is a computational-intensive benchmark that contains 80 operators of complex algorithms such as Wavelet decomposition (see Table~\ref{tab:evaluation-overall}), while Sense is a network-intensive application whose computations are simple (e.g., average).
Hence, the large $\alpha_{EEG}^*$ and small $\alpha_{Sense}^*$ are reasonable because the computation of EEG is important while network is vital for Sense.
2) \textit{Influence of device characteristics}.
The $\alpha_{EEG}^*$ changes from small in Fig.~\ref{fig:overall-results-energy}(a) to big in Fig.~\ref{fig:overall-results-energy}(b).
This is because the network change from Zigbee to WiFi significantly reduces the inter-block transmission time of EEG, which makes the computation the dominant factor in the optimization.
Unlike Wishbone, EdgeProg provides optimization goals with clear physical meaning which stay unchanged whatever the optimization task is, which makes EdgeProg more useful in practice.

\subsection{Overhead}

\textbf{Dissemination Overhead.}
The dynamic linkable and loadable binary sizes of the macro-benchmarks on three platforms: TelosB (TI MSP430), MicaZ (AVR ATMega128) and Raspberry Pi 3B+ (ARM Cortex-A53) supported by EdgeProg is summarized in Table \ref{tab:dissemination}.
\begin{table}
	\centering
	\caption{Dissemination size among platforms (Byte).}
	\iftechreport\else
	\vspace{-1em}
	\fi
	\label{tab:dissemination}
		\begin{tabular}{|c|c|c|c|c|}
		\hline
		\textbf{App.}	& \textbf{TelosB} & \textbf{MicaZ} & \textbf{Raspberry Pi}\\ \hline
		Sense & 4344 	& 6384 	& 4004 	\\ \hline
		MNSVG & 2756  	& 3460 	& 2280 	\\ \hline
		EEG   & 4500 	& 6276 	& 3920	\\ \hline
		SHOW  & 22952 	& 28660	& 14540 \\ \hline
		Voice & 32076 	& 42416 & 19336 \\
		\hline
	\end{tabular}
	\vspace{-2em}
\end{table}

We can see from the data that the binary size of SHOW and Voice is much bigger than other benchmarks, which is mainly due to the complexity of the algorithms they adopted such as FFT, MFCC.
Nevertheless, EEG has a smaller size compared with its large number of operators, which is mainly due to each of its tunnels shares the same procedures, and each procedure mainly contains one algorithm, wavelet decomposition, with different parameters.

\textbf{Run-time efficiency.}
\begin{figure*}[htbp]
	\centering
	\subfigure[Compare with virtual machines]{
		\begin{minipage}[t]{0.5\linewidth}
			\centering
			\includegraphics[width=.9\linewidth]{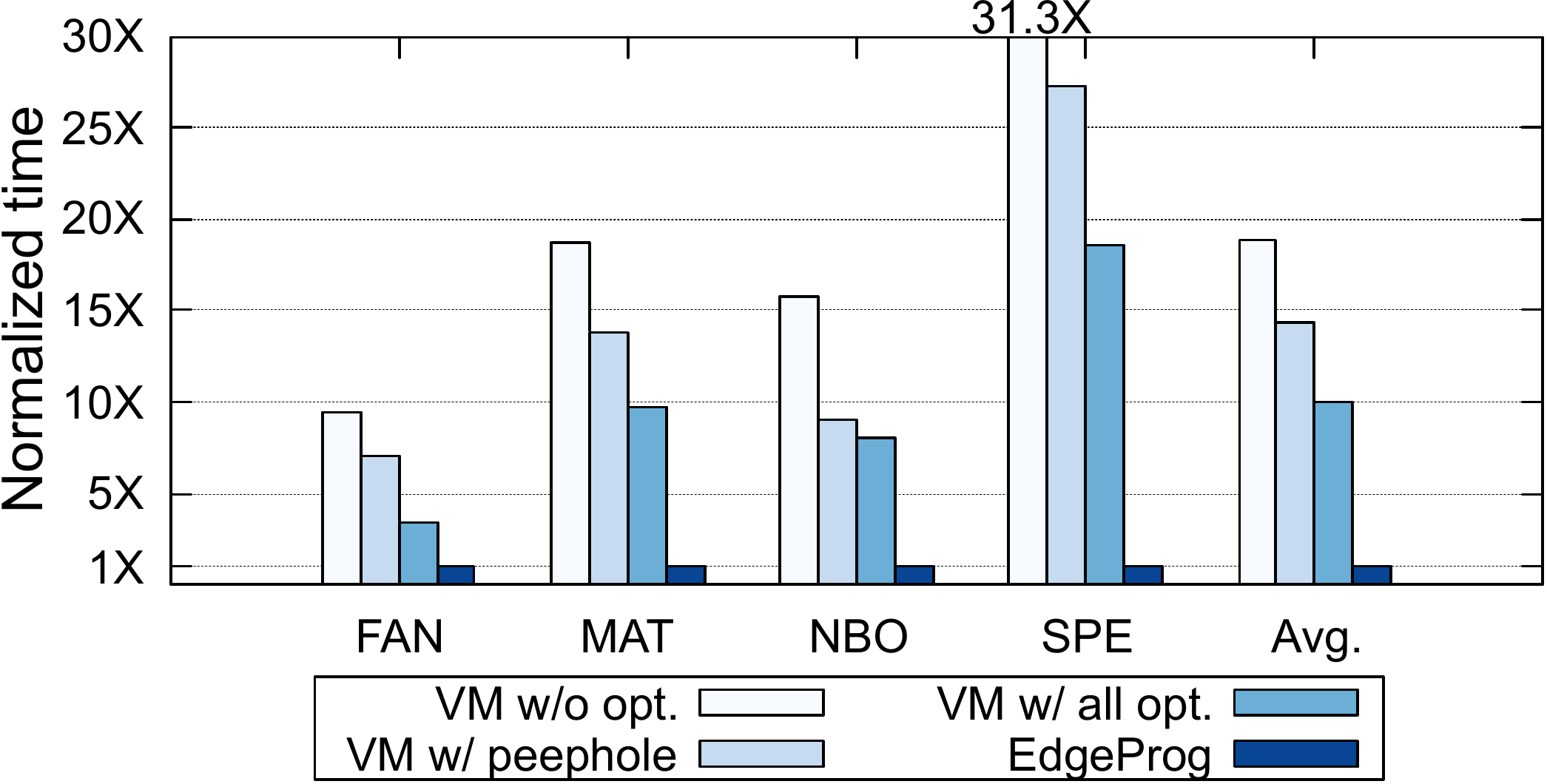}
		\end{minipage}%
	}%
	\subfigure[Compare with scripting languages]{
		\begin{minipage}[t]{0.5\linewidth}
			\centering
			\includegraphics[width=.9\linewidth]{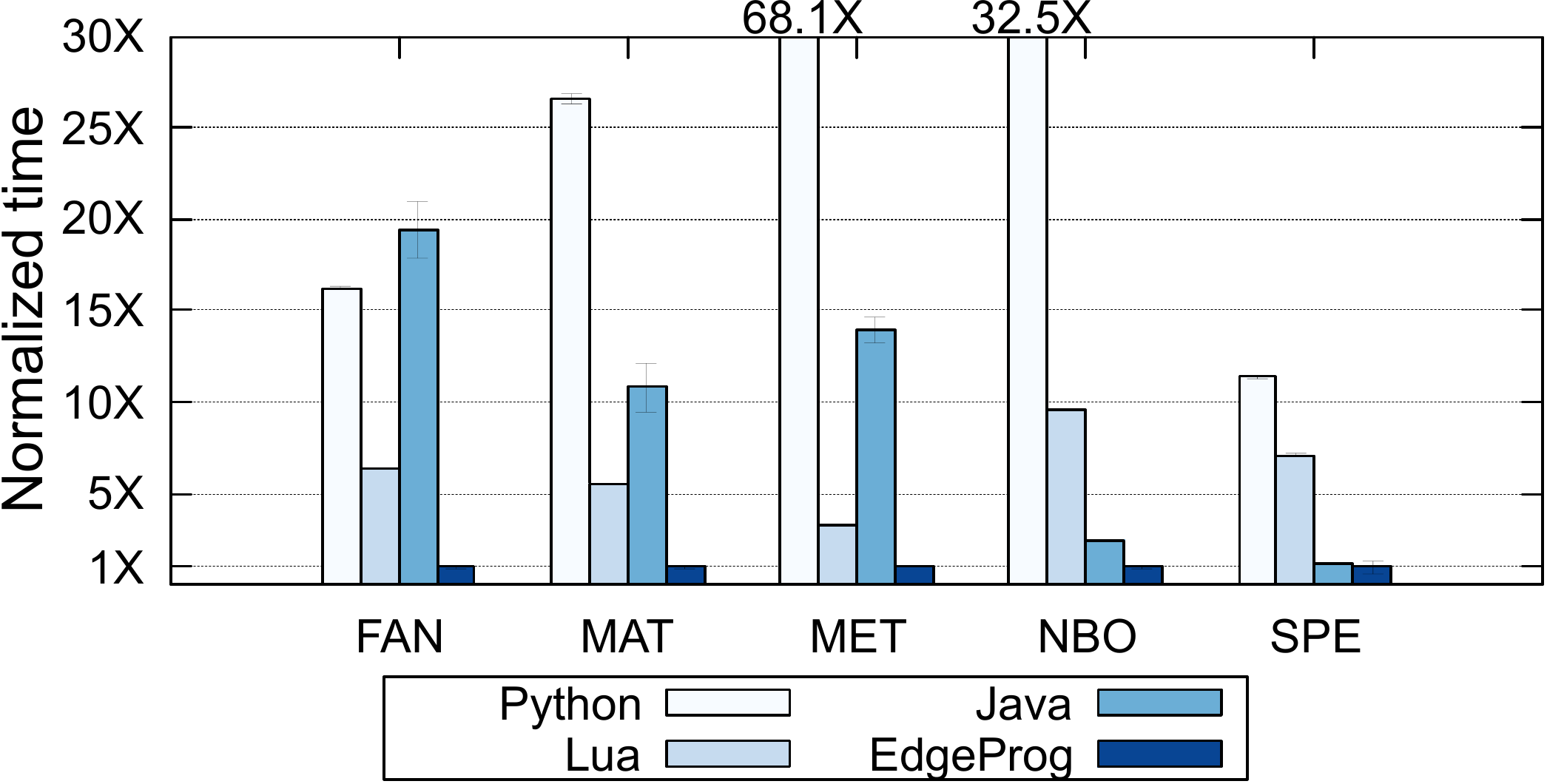}
		\end{minipage}%
	}%
	\centering
	\vspace{-1em}
	\caption{Run-time efficiency comparison between EdgeProg and design alternatives.}
	\vspace{-1em}
	\label{fig:vm-and-script}
\end{figure*}
In this section, we compare the run-time efficiency of the dynamic linking and loading technique with its alternatives: virtual machines (VMs) and scripting languages.
To eliminate the inherited overhead brought by different implementations, we use five micro-benchmarks from Computer Language Benchmark Game (CLBG).
CLBG is a language benchmark suite maintained by the Debian community.
The five benchmarks we excerpted are
Fannkuch problem (FAN), 
Matrix multiplication (MAT), 
Meteor predicting (MET), 
N-Body solution (NBO), 
and Spectral-Norm calculating (SPE).
We use CapeVM~\cite{reijers2018capevm}, a state-of-the-art Java VM developed for lightweight execution on embedded devices, as the representative of the VM technique.
CapeVM proposes various optimization strategies to accommodate different applications, and we set up the experiment with three settings: no optimization, only peephole optimization and all optimizations.
Moreover, we choose two scripting languages: Python (for popular) and Lua (for lightweight) along with Java, which is used in CapeVM, as our design alternatives of scripting languages.

Fig.~\ref{fig:vm-and-script} illustrates the experiment result.
Due to CapeVM do not support multidimensional arrays and floating points, the MET benchmark could not be implemented with CapeVM.
As shown in Fig.~\ref{fig:vm-and-script}(a), the VM method introduces a massive loss of run-time efficiency.
VM costs more than EdgeProg when executing the same benchmark by 9.98$\times$ on average and up to 31.32$\times$.
As for scripting languages and native Java illustrated in Fig.~\ref{fig:vm-and-script}(b), EdgeProg's dynamic linking and loading technique still outperforms than alternatives.
Python incurs the most overhead averaged 30.96$\times$ and Lua, being famous for its lightweight, still slows by 6.37$\times$ than ours. 
%



\subsection{Programming Language}
\begin{figure}[tbp]
	\centering
	\includegraphics[width=.8\linewidth]{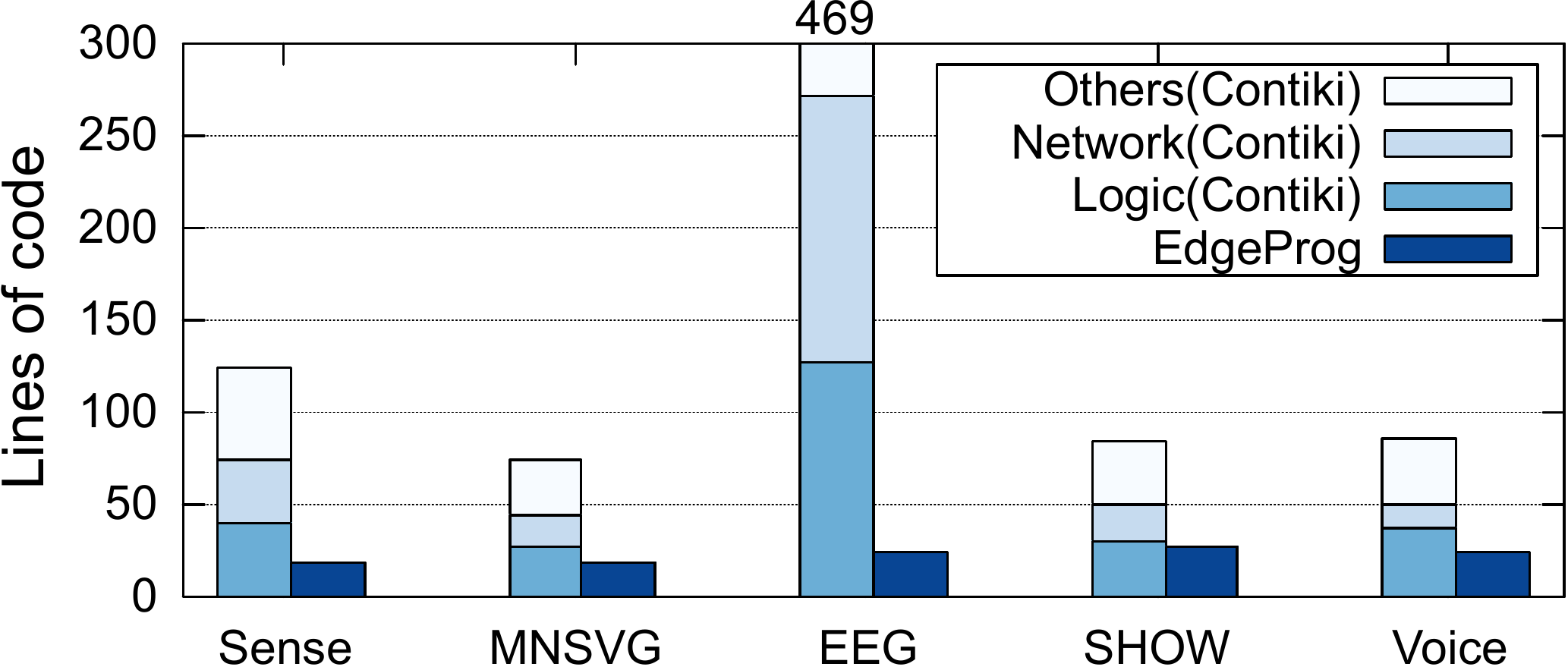}
	\vspace{-1em}
	\caption{Lines of code comparison between Contiki and EdgeProg. The "Logic", "Network" and "Others" represent the lines of code for expressing core application logic, inter-device network and others such as definition and included headers in Contiki source code.}
	\vspace{-1em}
	\label{fig:lines-of-code}
\end{figure}
In order to compare the reduction of lines of code via EdgeProg, we compare the lines of code of the macro-benchmarks described in Section~\ref{sec:evaluation-experiment-setup} written in traditional Contiki-style and EdgeProg-style.
Fig.~\ref{fig:lines-of-code} illustrates the comparison results.
Note that due to EdgeProg provides several data processing algorithms in advance to simplify the development procedure, we omit the lines of code for implementing the algorithms in Contiki-syle source code to achieve fair comparison and focus more on how EdgeProg helps for complex device interactions.
We can observe that 
(1) EdgeProg reduces the lines of code by 79.41\% on average.
This is because EdgeProg relieves users of writing complex inter-device interactions and other grammar necessaries.
Moreover, the virtual sensor and IFTTT abstraction contribute to the lines of code reduction for application logic. 
(2) EdgeProg reduces the development complexity, especially for applications with more devices.  
For example, the 80 stages of EEG application consists of 10 EEG devices, and each device owns eight stages.
Programming ten devices increases the lines of code multiple times.
While the relatively low reduction percentage of MNSVG (75.68\%), SHOW (67.86\%) and Voice (72.94\%) applications are partly because they need only one device and an edge device.

\subsection{Profiling Accuracy}
\label{sec:profiling-accuracy}
The correctness and accuracy of EdgeProg's latency-effective partition depend on the profiling method.
In this subsection, we evaluate the accuracy of profiling methods for both high- (\textit{e.g.,} Raspberry Pi) and low-end (\textit{e.g.,} TelosB) devices that we employ in EdgeProg. 

\begin{figure}[tbp]
	\centering
	\includegraphics[width=.8\linewidth]{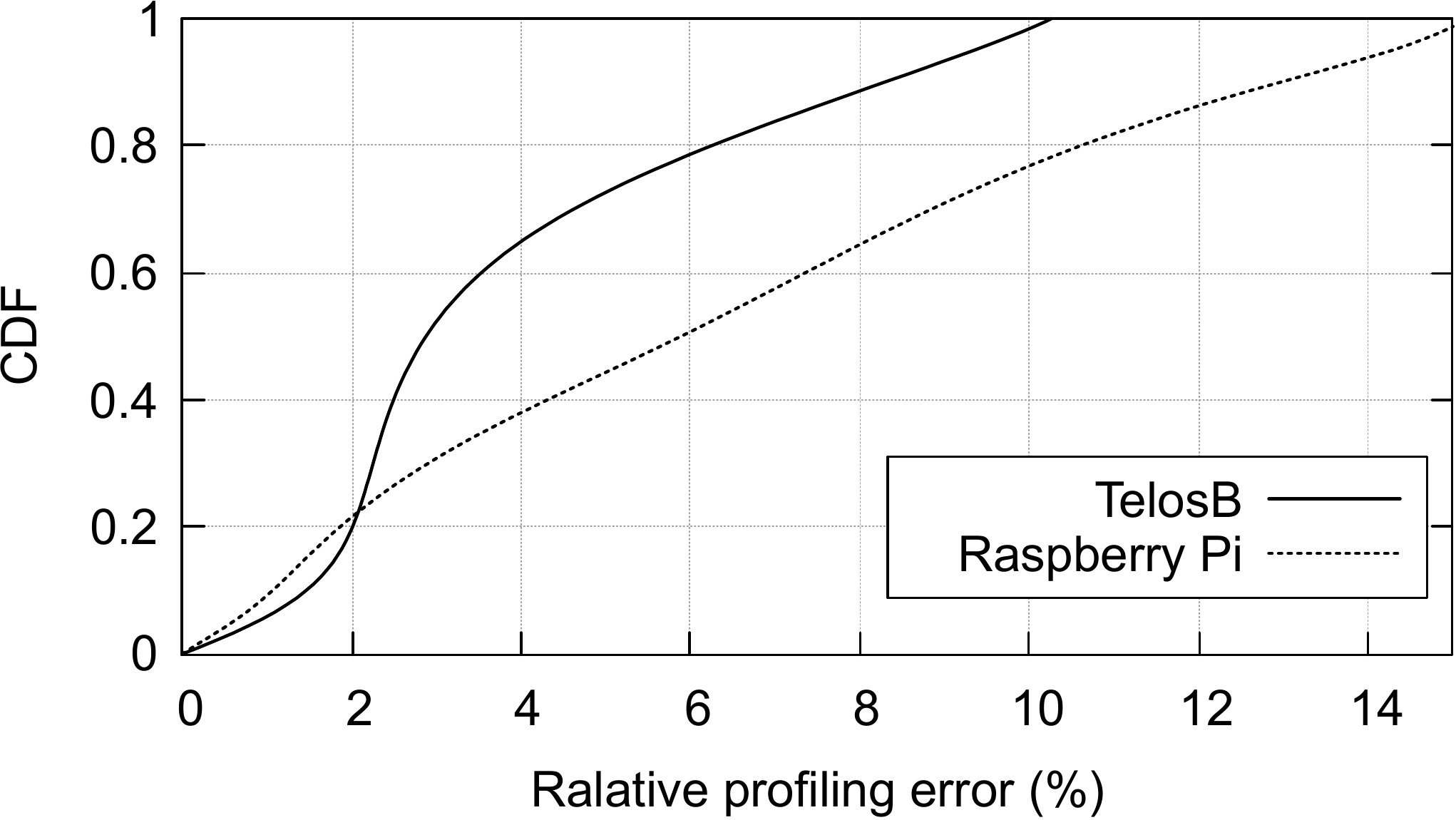}
	\vspace{-1em}
	\caption{Profiling accuracy of high-/low-end devices.}
	\vspace{-1em}
	\label{fig:profile-accuracy}
	
\end{figure}

We use \texttt{mspsim} to profile the applications of TelosB, and a near cycle-accurate simulator \texttt{gem5} for modern platforms such as Raspberry Pi. 
For \texttt{gem5}, we use the system call emulation (SE) mode with the compiled binary as input to avoid the additional overhead of its full-system mode.
The results are shown in Fig.~\ref{fig:profile-accuracy}.
\texttt{mspsim} could achieve 90\%+ accuracy over 97.6\% of test cases.
Nevertheless, only 87.1\% cases of \texttt{gem5} reach 90\%+ accuracy, which is mainly due to the frequency fluctuation of CPUs and background processes of Raspberry Pi.

%
\section{Discussion}
\label{sec:discussion}
In this section, we discuss several important open issues of EdgeProg.

\textbf{Dynamic evolving scenario of EdgeProg.}
Partitioning the application is not a one-shot job in real-world deployments.
The optimal partition may change due to the disturbance of wireless network or even device breakdown.

EdgeProg supports the dynamic partition update during run-time.
The environmental variation is captured by our network profiler deployed on the edge device (Section~\ref{sec:architecture}).
EdgeProg periodically checks if the environmental variation leads to sub-optimal performance for a certain length of time (i.e., tolerance time), EdgeProg will start the partition updating process.
The updating process includes compiling the platform-specific binary and dispatching binaries to devices for reprogramming.
Users could adjust the environmental sensitivity of EdgeProg by setting the appropriate tolerance time in case of frequent updating that brings high reprogramming overhead.

\begin{figure}[tbp]
	\centering
	\includegraphics[width=.8\linewidth]{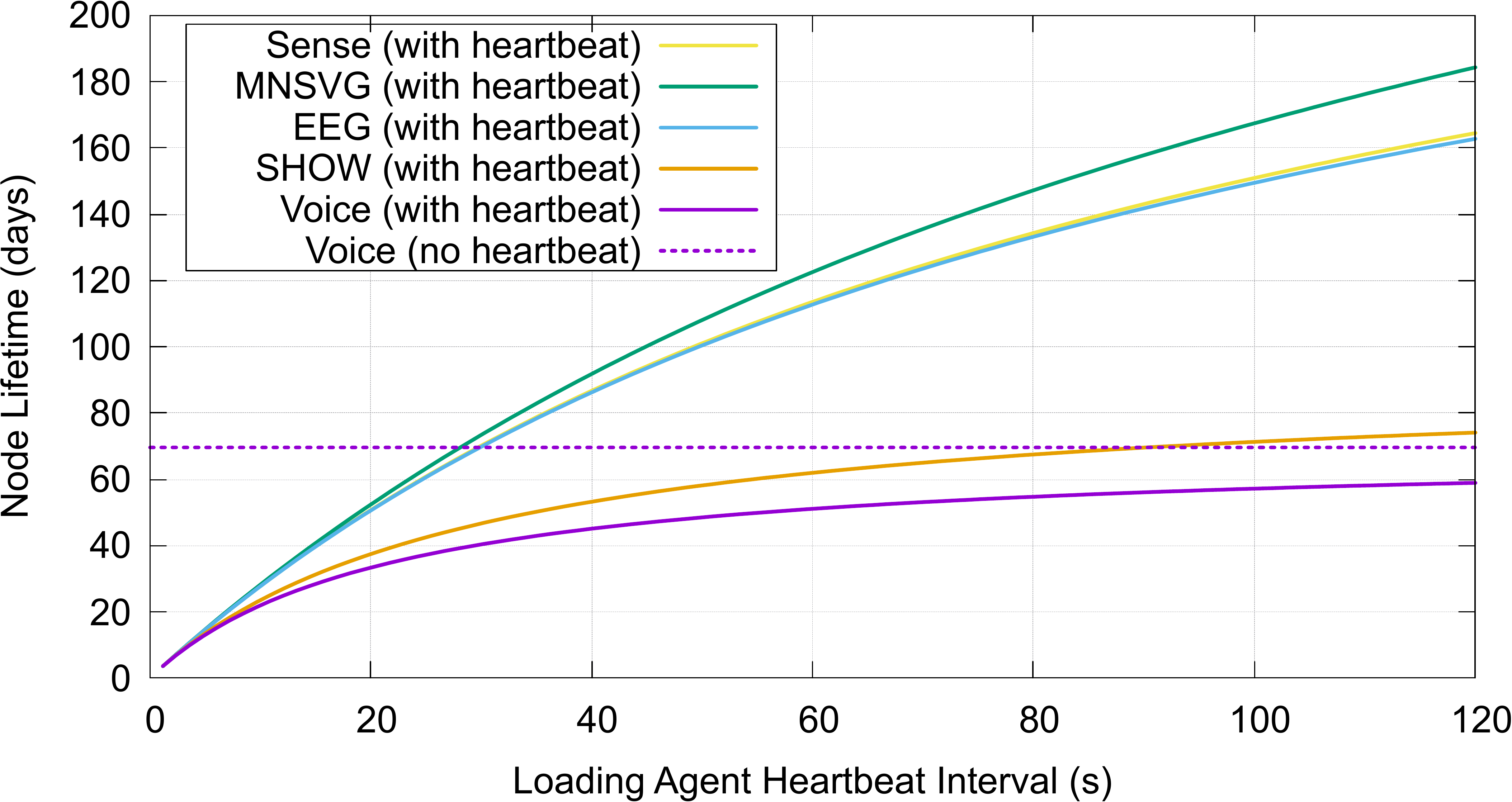}
	\vspace{-1em}
	\caption{Relationship between node lifetime against heartbeat interval.}
	\vspace{-0.5em}
	\label{fig:loading-agent-energy}
	\vspace{-1em}
\end{figure}

\textbf{Energy drain of the loading agent.}
The loading agent in EdgeProg communicates with the edge server periodically (i.e., heartbeat) to check if there is a new binary to load and execute. 
When there is a new binary, the loading agent downloads the binary and loads it.
Hence, the energy consumption of the loading agent is two-fold: periodical heartbeat and binary load. 
Inspired by~\cite{dong2015optimizing}, we build an analytical model and illustrate the energy impact of the loading agent in Fig.~\ref{fig:loading-agent-energy}.
\iftechreport
We take TelosB node in EdgeProg as an example to illustrate the overhead of the loading agent.
Inspired by~\cite{dong2015optimizing}, we formulate the numerical relationship of node lifetime against loading agent heartbeat interval as:
\begin{equation}
	L(t_{hb}) = \frac{UB}{\frac{f(P_{radio}+P_{MCU})t+E_{heartbeat}(t_{hb})+E_{load}}{t+t_ps_p}+rUB}
\end{equation}

$L(t_{hb})$ denotes the lifetime against the heartbeat period $t_{hb}$. 
$U$ denotes the operation voltage. 
$B$ denotes the battery capacity (set to 2200mAh for NiMH battery). 
$f$ denotes the duty-cycle of the running application (set to 0.1\% according to~\cite{musaloiu2008koala}). 
$P_{radio}$ and $P_{mcu}$ denote the average power of radio and MCU. 
$t$ denotes how often a new binary is disseminated (set to 10 days for illustration). 
$E_{heartbeat}$ and $E_{load}$ denote the energy consumption of heartbeats and binary loading. 
$t_p$ stands for the average time to receive one byte of binary and $s_p$ means the binary size. 
$r$ is the self-discharge rate per day (set to $\frac{0.33}{365}$ because we assume the batteries lose one third after a year).

Based on the above analytical model, we could illustrate the energy impact of the loading agent in Fig.~\ref{fig:loading-agent-energy}.
\else
We leave the model formulation and to our technical report~\cite{li2021edgeprogtechreport} due to the limited space.
\fi
We set the battery capacity to 2200mAh, and assume new binaries are generated every ten days.
We can see from the figure that the heartbeats indeed affect the battery lifetime. 
The loading agent leads to a 14.5\% and 26.1\% decrease of Voice benchmark when the heartbeat interval is 120s and 60s, respectively. 
Hence, considering the tradeoff between timeliness of binary loading and the energy drain, we set the heartbeat interval to 60s in our implementation by default, and we allow users to modify the interval to meet their individual needs.

\textbf{Limitations of EdgeProg language.}
%
Although EdgeProg language shows its simplicity and expressiveness in our evaluation, it still owns two limitations:

(1) Limited support for implementing new peripheral libraries.
The software library (e.g., the Light\_Solar library of TelosB in line~\ref{loc:programming-language-telosb-library} and GMM library in line~\ref{loc:programming-language-vsensor-library} of Fig.~\ref{fig:programming-language}) is one of the key components of EdgeProg's language and block-based data flow representation.
In the current implementation of EdgeProg, the common peripheral libraries and 17 data processing algorithms for virtual sensors on four platforms (TelosB, MicaZ, RPI and PC) are included in EdgeProg as we noticed in Section~\ref{sec:programming-language}.
We are working on the porting tutorials and code templates for implementing new libraries for peripherals and platforms.
(2) Algorithms with feedback.
The applications written with EdgeProg language finally transformed into a DAG for optimization, which introduces a restriction that the algorithms with feedback could not be expressed by EdgeProg language.
The rationale behind this limitation is the feedback may incur directed cyclic graph (DCG), which is not solvable under our current formulation.
We do not take the DCG into consideration temporarily is because: \textit{a)} the DAG representation is widely adopted by the state-of-the-art distributed computation systems (e.g., Apache Storm), and \textit{b)} considering DCG would incur additional solving overhead, which will affect the run-time efficiency of EdgeProg. 
We consider generalizing the formulation of EdgeProg to support more kinds of application topologies as our future work.

\textbf{Time and energy profiling.}
In EdgeProg, we gather the time and energy information of each logic block by leveraging cycle-accurate simulators.
This method achieves fair accuracy (results shown in Fig.~\ref{fig:profile-accuracy}) when the hardware parameters (e.g., frequency and existing workload of MCU) of the deployment environment is similar to the simulation, which is common in IoT scenario because the devices mostly operate under a certain performance level~\cite{li2020automatic}.

Nevertheless, the auto-scaling technique in the modern high-end edge server brings hundreds or thousands of performance levels, which makes it painful to profile the performance under each hardware parameter setting.
Towards this situation, we consider generating the full profile with incomplete data through the efficient learning-driven prediction algorithm~\cite{hu2019learning}.
Moreover, we also consider integrating the time estimation against different current workloads proposed in~\cite{li2021queec} on edge devices to further improve the profiling accuracy of EdgeProg.

\vspace{-1em}
\section{Related Work}
\label{sec:related-work}
EdgeProg borrows heavily from existing works. In the following paragraphs,
We discuss three main categories:
IoT application programming, 
code partitioning and offloading,
as well as edge computing. 

\textbf{IoT application programming.}
The traditional approach for IoT programming is device-centric~\cite{guan2017tinylink}
, i.e., the application logic resides on the IoT devices. For example, developers may write application-specific sensor data processing or multi-hop forwarding based on IoT operating systems such as TinyOS or Contiki OS. 

To simplify application programming for multi-device interaction, developers can adopt trigger-action programming like IFTTT on edge/cloud servers so that the whole app logic resides on the server.
The IoT nodes perform general functions like sensor data sampling and data transmissions. 
IFTTT programming is widely adopted in the industry, such as Samsung SmartThings and Microsoft Flow. 
It also attracts a lot of research attention from academia~\cite{heo2017rt,ur2014practical}. 
For example, a recent work, RT-IFTTT~\cite{heo2017rt}, enhances the traditional IFTTT syntax. RT-IFTTT’s key idea is to dynamically adjust the sensor data polling intervals to satisfy both energy and real-time constraints. 
EdgeProg inherits from IFTTT’s server-centric programming model but differs from existing works in two important ways. First, we enhance the IFTTT syntax with special consideration on data-intensive computation. Second, we enable much more flexible server-device cooperation by supporting code partitioning and dynamic code loading on the device, compared with RT-IFTTT which only supports adjusting data sampling intervals.  

In retrospect, a similar work to ours is Tenet~\cite{gnawali2006tenet} in the sensor network literature. Tenet assumes a two-tier network architecture consisting of ordinary sensor nodes and master nodes. Tenet’s principle is to place the application-specific logic on the master tier using a dataflow program. 
The master nodes can dynamically task sensor nodes to process data locally. 
In EdgeProg, the edge server plays an equivalent role to the master nodes. EdgeProg differs from Tenet in the language design, 
device-side system support, 
and performance optimizations.

\textbf{Code partitioning and offloading.}
Code offloading to heterogeneous IoT nodes needs system support at the device-side. 
A virtual machine is a common approach to mask heterogeneity. 
There is rich literature in designing flexible and efficient VMs on resource-constrained nodes, including Mate~\cite{levis2002mate}, CapeVM~\cite{reijers2018capevm}, JVM, \textit{etc}.
In addition, a large number of offloading algorithms builds on top of VMs, e.g., Tenet~\cite{gnawali2006tenet}, ASVM~\cite{levis2005active}.
Besides VM, there are other more lightweight approaches such as Linux containers, RPC~\cite{guan2019queec}, loadable modules~\cite{cao2016tinysdm}.
We adopt the loadable module approach in EdgeProg. This is because execution efficiency is critical for energy-constrained IoT nodes and native code runs much faster than VM instructions~\cite{dunkels2006run,dong2009dynamic}.

There is rich literature in code partitioning and offloading algorithms for performance optimizations.
LEO~\cite{georgiev2016leo} presents an offloading algorithm targeting mobile sensing applications. 
LEO makes use of domain specific signal processing knowledge to smartly distribute the sensor processing tasks across the broader range of heterogeneous computational resources of high-end phones (CPU, co-processor, GPU and the cloud).
LEO achieves fine-grained energy control by exposing internal pipeline stages to the scheduler.
Queec~\cite{guan2019queec} takes the user-perceived quality of experience (QoE) into offloading decision and makes efforts to achieve the lowest latency.
EdgeProg shares similarities with many existing algorithms to optimize performance metrics such as latency or energy. 
However, EdgeProg uses a different formulation considering multiple rules execution, cached values, and concurrent execution on different IoT nodes. 

Also, there are a variety of efforts concerning the partitioning and deployment of DAG-represented applications.
Wishbone~\cite{newton2009wishbone} presents a code partitioning algorithm among resource-constrained sensor nodes and the server to process data-intensive applications by cutting the unnecessary edges of the DAG.
P-EDF-omp~\cite{wang2020partitioning} proposes a DAG partitioning algorithm to schedule tasks across multiple processors while keeping the hard real-time guarantee for OpenMP applications.
Moreover, Storm is a widely used analytic framework, and its application is constructed using DAG.
HeteroEdge~\cite{zhang2019hetero} focuses on partitioning and scheduling the Storm applications between CPUs and GPUs to achieve better latency.
Different from the above literature, the DAG partitioning algorithm of EdgeProg considers the weights on both vertices and edges, which further leads to a different formulation and solution.

\textbf{Edge computing systems.}
EdgeProg runs on existing edge platforms and focuses on programming IoT nodes connected to the edge.
Most existing work of edge computing focuses on how to program the edge itself. 
In ParaDrop~\cite{liu2016paradrop}, the edge service deployment is initiated and controlled by a cloud server.
ParaDrop employs the container technology for the concurrency and isolation between edge services.

As for programming the edge-connected nodes, EveryLite~\cite{li2018everylite} proposes a lightweight scripting language (37KB of core runtime size) extended from Lua for developers to build applications.
Nevertheless, EdgeProg chooses the native C approach to further reduce the run-time overhead.
Furthermore, EveryLight only focuses on programming one node, while EdgeProg also takes the edge device and connected nodes into consideration.
Considering the coordinated programming for both the edge and nodes, the most similar and recent work is DDFlow~\cite{noor2019ddflow}.
Its idea borrows from the existing macroprogramming approach~\cite{gummadi2005macro}
, which aim to build applications in the whole network point-of-view (POV) rather than per-node POV.
DDFlow presents a visual programming interface for developers to state their application as a task graph.
EdgeProg employes a more declarative way with a domain-specific language rather than graphical programming, and achieves the lowest latency even in the multi-rule situation while DDFlow only considers optimizing one application per time.

\section{Conclusion}
\label{sec:conclusion}

This paper presents EdgeProg, an edge-centric programming system with automatic code partitioning.
In EdgeProg, we provide developers, especially non-experts, with an easy-to-use yet expressive programming language.
Build upon the global view of our language, the code partitioner finds the best placement for each part of the application through an ILP formulation, which could be efficient and optimally solved.
The key insight is that we make the best use of the computation ability of each device to achieve better performance.
Evaluations show that EdgeProg could reduce the task execution latency by 31.65\% for ZigBee networks and 10.26\% for WiFi networks.
For energy, EdgeProg saves 14.8\% and 40.8\% on average compared with state-of-the-arts.
Also, EdgeProg reduces the lines of code by 79.41\%.

\iftechreport
\begin{appendices}

\section{Examples of EdgeProg Programming Language}
\label{sec:appendix-language}
In this appendix section, we illustrate five examples from real-world projects and research efforts that could be expressed using EdgeProg programming language.

\textbf{RFace.}
Xu et al. propose RFace~\cite{xu2021rface}, a facial anti-spoofing and privacy-preserving authentication system with COTS RFID devices.
RFace leverages an RFID tag and a reader to measure Received Signal Strength (RSS) and phase of RF signals.
Furthermore, RFace first prepossesses the RFID data and extracts 3D facial geometry and biomaterial features for spooﬁng attack resistance.
We illustrate the example code to express the application logic of the authentication stage of RFace in Fig.~\ref{fig:appendix-pl-rface}.

\begin{figure}
	\centering
	\begin{lstlisting}
Application RFace{
	Configuration{
		RPI A(RSS, RFPhase);
		Arduino B(Alarm);
		Edge E();
	}
	Implementation{
		VSensor Preprocess("DN, REA"){
			preprocess.setInput(A.RSS, A.RFPhase);
			DN.setModel("Denoise");
			F1.setModel("Rearrange");
			Preprocess.setOutput(<pre_t>);
		}
		VSensor DDDS("DI, CAL, COS"){
			DDDS.setInput(Preprocess);
			DI.setModel("Division");
			CAL.setModel("Calculation");
			COS.setModel("CosineSimilarity")
			COS.setOutput(<string_t>, "fake");
		}
	}
	Rule{
		IF(DDDS=="fake")
		THEN(B.Alarm);
	}
}
	\end{lstlisting}
	\vspace{-1em}
	\caption{Code snippets of the \texttt{RFace} application.}
	\label{fig:appendix-pl-rface}
\end{figure}

\textbf{LimbMotion.}
LimbMotion~\cite{zhou2019limbmotion} is a recent research project of Zhou et al. to track the 3D posture using a smartwatch.
The 3D posture of a limb is defined by the relative positions among main joints, e.g., shoulder, elbow, and wrist for an arm.
LimbMotion leverages both the inertial measurement units (IMUs) sensor and the acoustic sensor to achieve fast and accurate limb posture estimation.
The acoustic signal is treated by several signal processing algorithms to obtain accurate distance information.
Then, LimbMotion leverages a two-step filtering approach that combines a complementary filter and a Kalman filter to process the IMU data.
The example code that illustrates the core application logic using the EdgeProg programming language is shown in Fig.~\ref{fig:appendix-pl-limbmotion}.

\begin{figure}
	\centering
	\begin{lstlisting}
Application LimbMotion{
	Configuration{
		RPI V(Voice);
		RPI I(IMU); (*@\label{loc:programming-language-telosb-library-1} @*)
		Edge E(Database);
	}
	Implementation{
		VSensor AcousticRanging("WS, F1, SD, F2"){
			AcousticRanging.setInput(V.Voice);
			WS.setModel("WindowSlicing");
			F1.setModel("FFT");
			SD.setModel("EnergyThreshold");
			F2.setModel("IFFT");
			AcousticRanging.setOutput(<vr_t>);
		}
		VSensor PointCloudEst("CF, KF"){
			PointCloudEst.setInput(I.IMU);
			CF.setModel("ComplementaryFilter");
			KF.setModel("KalmanFilter");
			PointCloudEst.setOutput(<pce_t>);
		}
		VSensor LimbPosition("LM"){
			LimbPosition.setInput(AcousticRanging, PointCloudEst);
			LM.setModel("LMAlgorithm");
			LimbPosition.setOutput(<float_t,float_t,float_t,float_t,float_t,float_t>);
		}
	}
	Rule{
		IF(LimbPosition[0]==LimbPosition[3]&&LimbPosition[1]==LimbPosition[4])
		THEN(E.Database("The leg is pointing!"));
	}
}
	\end{lstlisting}
	\vspace{-1em}
	\caption{Code snippets of the \texttt{LimbMotion} application.}
	\label{fig:appendix-pl-limbmotion}
\end{figure}

\textbf{RepetiveCount.}
Zhang et al. advocate RepetiveCount~\cite{zhang2021repetitive} for repetitive activity counting in videos.
This work incorporates the corresponding sound along with the video data into the repetition counting process.
RepetiveCount takes video clips as inputs and outputs the counting result for each clip by adopting a 3D convolutional network.
For the sound stream, RepetiveCount relies on a 2D convolutional network, which takes the sound spectrogram generated by the short-time Fourier transform as input and outputs the counting result in the same way as the sight stream.
Finally, a reliability estimation module based on several fully connected neural network decides what prediction to use.
In Fig.~\ref{fig:appendix-pl-repetivecount}, we illustrate the application logic of RepetiveCount using the EdgeProg programming language.

\begin{figure}
	\centering
	\begin{lstlisting}
Application RepetiveCount{
Configuration{
	RPI A(Video);
	RPI B(Voice);
	Edge E(Database);}
Implementation{
	VSensor ResBlock3D("CNN1, RB3D"){
		ResBlock3D.setInput(A.Video);
		CNN1.setModel("CNN", "VideoCNN.pt");
		RB3D.setModel("ResBlock3D", "RB3D.pt");
		AcousticRanging.setOutput(<rb3d_t>);
	}
	VSensor ResBlock2D("FFT, CNN2, RB2D"){
		ResBlock3D.setInput(B.Voice);
		FFT.setModel("FFT");
		CNN2.setModel("CNN", "VoiceCNN.pt");
		RB2D.setModel("ResBlock2D", "RB2D.pt");
		AcousticRanging.setOutput(<rb2d_t>);
	}
	VSensor RepeCtVideo("{FCV1_1, FCV1_2}, SUMV1"){
		RepeCtVideo.setInput(ResBlock3D);
		FCV1_1.setModel("FC", "FCV1.pt");
		FCV1_2.setModel("FC", "FCV1.pt");
		SUMV1.setModel("Sum", "FCV1_1", "FCV1_2");
		RepeCtVideo.setOutput(<RepeCtVideo_t>);
	}
	VSensor RepeCtVoice("{FCV2_1, FCV2_2}, SUMV2"){
		RepeCtVoice.setInput(ResBlock2D);
		FCV2_1.setModel("FC", "FCV2.pt");
		FCV2_2.setModel("FC", "FCV2.pt");
		SUMV1.setModel("Sum", "FCV2_1", "FCV2_2");
		RepeCtVoice.setOutput(<RepeCtVoice_t>);
	}
	VSensor RelEst("FCV3"){
		RelEst.setInput(ResBlock2D);
		FCV3.setModel("FC", "FCV3.pt");
		RelEst.setOutput(<re_t>);
	}
	VSensor CountPredict("CONCAT, MUL"){
		CountPredict.setInput(RepeCtVideo, RepeCtVoice, RelEst);
		CONCAT.setModel("VecConcat", "RepeCtVideo", "RepeCtVoice");
		MUL.setModel("MatMul", "CONCAT", "RelEst")
		CountPredict.setOutput(<float_t>);
	}}
Rule{
	IF(SUM+=CountPredict>=1)
	THEN(E.Database("UPDATE ct SET ClothCt={SUM} WHERE ID=100")&&E(SUM=0));}
}
	\end{lstlisting}
	\vspace{-1em}
	\caption{Code snippets of the \texttt{RepetiveCount} application.}
	\label{fig:appendix-pl-repetivecount}
\end{figure}

\textbf{Hyduino.}
Hyduino~\cite{web-dfrobot-hyduino} is a real-world IoT project we excepted from DFRobot.com.
The design purpose of Hyduino is to monitor the pH value, temperature, soil humidity of a plant.
If the environmental values exceed or below the appropriate scale, Hyduino will perform the corresponding action to keep a suitable value.
Fig.~\ref{fig:appendix-pl-hyduino} illustrates the EdgeProg application of Hyduino.

\begin{figure}
	\centering
	\begin{lstlisting}
Application Hyduino{
	Configuration{
		Arduino A(PH);
		Arduino B(TEMPERATURE, HUMIDITY);
		Arduino C(turnOnFAN);
		Arduino D(openPHPump);
		Arduino E(SDCardWrite);
		Edge E(LCD_SHOW);
	}
	Rule{
		IF(A.PH>7.5 && B.TEMPERATURE>28 && B.HUMIDITY>44)
		THEN(C.turnOnFAN && D.openPHPump && SDCardWrite("Start") && E.LCD_SHOW("PH: %f, Temp: %f", A.PH, B.TEMPERATURE));
	}
}
	\end{lstlisting}
	\vspace{-1em}
	\caption{Code snippets of the \texttt{Hyduino} application.}
	\label{fig:appendix-pl-hyduino}
\end{figure}

\textbf{SmartChair.}
Error sitting posture is harmful to the cervical vertebra of a person.
SmartChair~\cite{web-dfrobot-smartchair} is a project that uses Ultrasonic and PIR sensor to monitor the sitting posture.
If the sitting posture is not standard, SmartChair leverages an alarm to notify the user to correct the posture.
We illustrate the SmartChair application using EdgeProg language in Fig.~\ref{fig:appendix-pl-smartchair}.

\begin{figure}
	\centering
	\begin{lstlisting}
Application SmartChair{
	Configuration{
		Arduino A(UltraSonic, PIR);
		Arduino B(Alarm);
		Edge E();
	}
	Implementation{
		VSensor US_Distance("PRE, CAL"){
			US_Distance.setInput(A.UltraSonic);
			PRE.setModel("US_PRE");
			F1.setModel("US_CAL_DIST");
			US_Distance.setOutput(<float_t>);
		}
	}
	Rule{
		IF(Distance>20 && Distance<3000 && A.PIR=1)
		THEN(B.Alarm);
	}
}
	\end{lstlisting}
	\vspace{-1em}
	\caption{Code snippets of the \texttt{SmartChair} application.}
	\label{fig:appendix-pl-smartchair}
\end{figure}

\section{Performance Comparison for Solving the QP and LP formulation}
\label{sec:appendix-gurobi}
In EdgeProg, the problem scale (i.e., the total number of $X_{b_is}$) is the product of the number of available devices and the logic blocks. 
In realistic scenarios, take the five benchmarks in our manuscript as examples.
The largest scale is the EEG application (80 blocks$\times$11 devices), the smallest scale is the MNSVG application (4 blocks$\times$3 devices). 
The average scale across the five examples is 211. 
Furthermore, we surveyed 50 IoT projects from several popular IoT communities such as hackster.io and DFRobot.com.
The average scale among the investigated projects is 100-150.

To compare the cost for solving the linear programming (LP) formulation and the original quadratic programming (QP) formulation, we conduct an experimental evaluation of the two methods. 
We use the Gurobi solver instead of \texttt{lp\_solve} that we used in EdgeProg. 
The usage of Gurobi is because it shows the best solving performance among the commonly used solvers [1] and it supports both LP and QP problems which leads to a fair comparison of the two formulations. 
The reason we use \texttt{lp\_solve} instead of Gurobi in the implementation of EdgeProg is that Gurobi is commercial software which may be an obstacle when we make EdgeProg open source for the community. 
We conduct experiments under the identical setting of the edge server to the one in our paper (2.8GHz i7-7700HQ CPU and 16GB memory).
We use the energy consumption formulation (Equ.~(\ref{equ:obj-origin-energy}) for QP and Equ.~(\ref{equ:obj-energy-final}) for LP in the paper) as an example. 
The experimental results are illustrated in Fig.~\ref{fig:gurobi-time-vs-scale} and \ref{fig:gurobi-stacked-time}.

\begin{figure}[t]
	\centering
	\includegraphics[width=.85\linewidth]{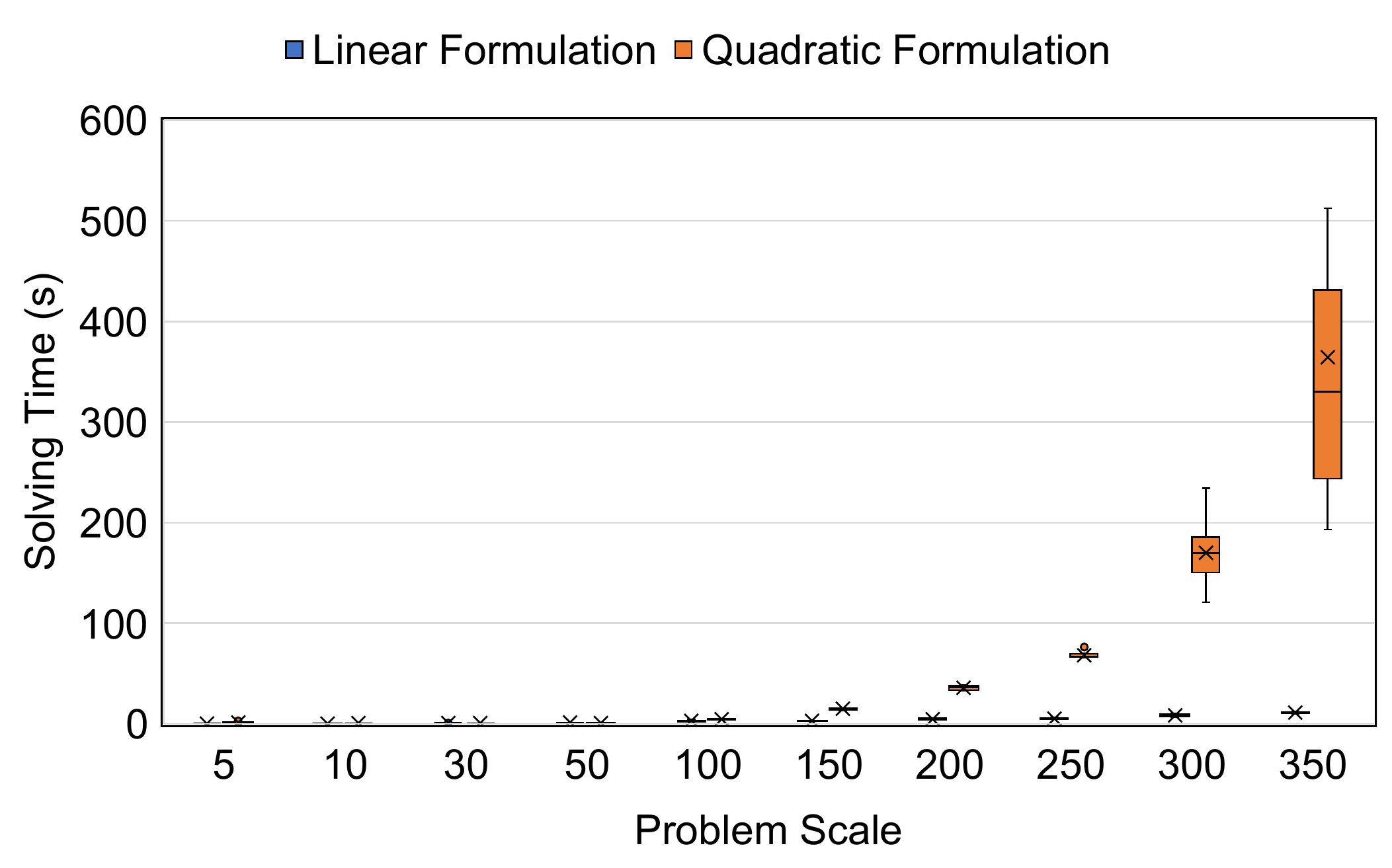}
	\vspace{-1em}
	\caption{Total solving time of LP and QP using Gurobi Solver.}
	\vspace{-1em}
	\label{fig:gurobi-time-vs-scale}
\end{figure}

\begin{figure}[t]
	\centering
	\includegraphics[width=.85\linewidth]{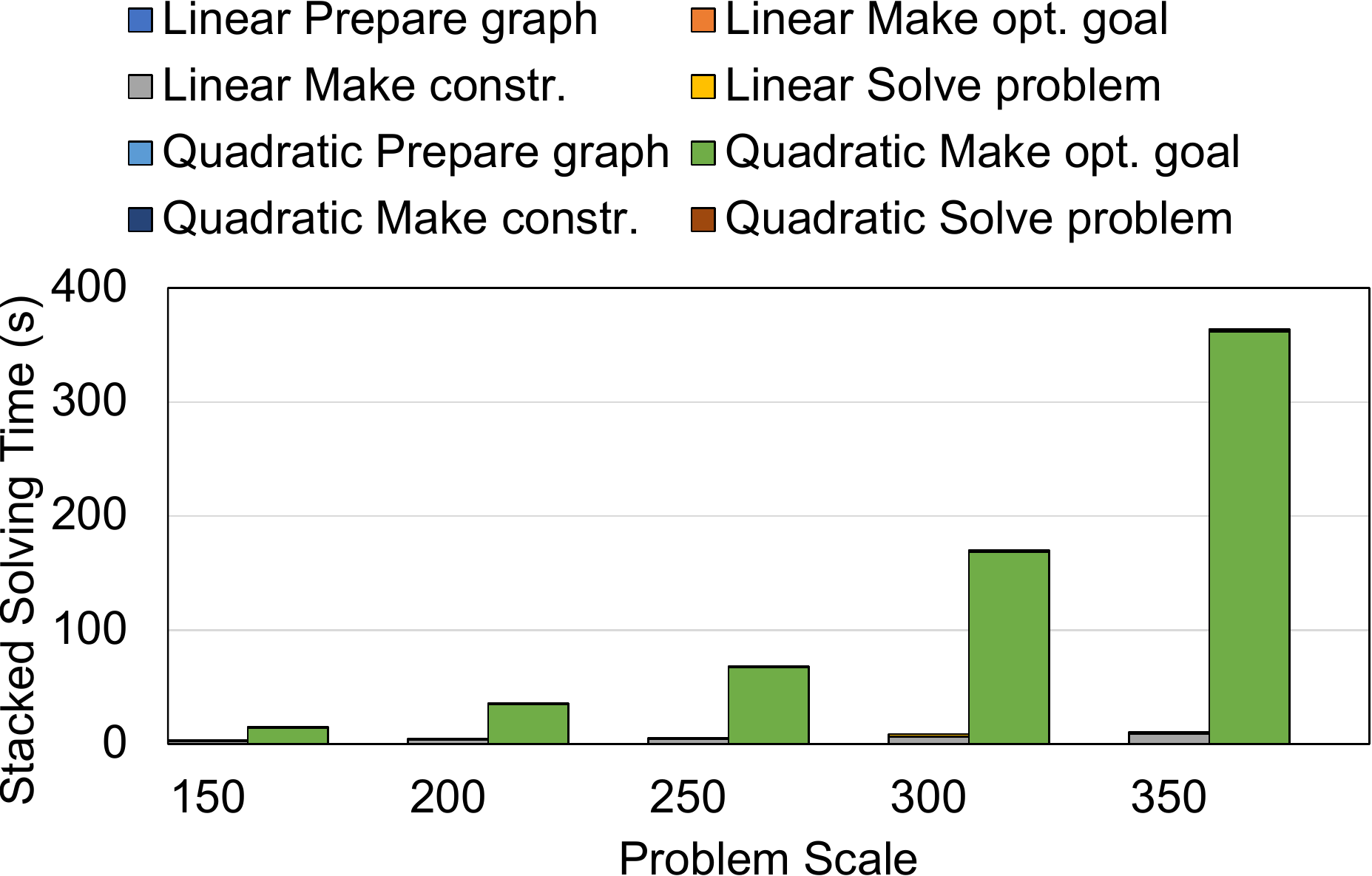}
	\vspace{-1em}
	\caption{Solving time breakdown of LP and QP.}
	\vspace{-1em}
	\label{fig:gurobi-stacked-time}
\end{figure}

The results in Fig.~\ref{fig:gurobi-time-vs-scale} show that the total solving time of QP increases much faster than that of LP. 
The EEG application (scale=811) is nearly unsolvable under the QP formulation. 
When the problem scale goes to the average problem size (scale$\sim$200), the solver needs 35.79s to solve the QP while only 4.89s for LP. 
To further investigate how the solving time increases, we break down the solving stages of LP and QP in Fig.~\ref{fig:gurobi-stacked-time}. 
We omitted the data of several problem scales to make the figure clear because the solving time among problem scales differs dramatically. 
The solving time includes all the stages that are needed to solve the problem, such as prepare the logic graph, constructing the optimization goal and constraints, and the solving time. 
We can see that for the LP problem, most of the time is consumed on making constraints. 
This is because the constraints of the LP problem are more than the QP’s, which increases linearly (four more constraints for a new variable). 
As for the QP formulation, much time is spent on making the optimization goal, which increases quadratically due to the size of $X_{b_is}X_{b_{i^{'}}s^{'}}$ matrix in the QP formulation (Equ.~(\ref{equ:obj-origin-energy}) in our paper) increases quadratically.


\end{appendices}

\fi

\bibliographystyle{IEEEtran}
\bibliography{reference/ref,reference/new_refs}
%
%
%

%
%
%
%
%

\iftechreport\else
\vspace{-2em}
\begin{IEEEbiography}[{\includegraphics[width=1in,height=1.25in,clip,keepaspectratio]{photo/Borui_Li}}]{Borui Li}
	received the B.S. degree in computer science from Nanjing University of Posts and Telecommunications, China, in 2017.
	He is now a Ph.D. candidate at Zhejiang University.
	His research interests include Internet of Things and edge computing systems  .
\end{IEEEbiography}

\vspace{-1em}
\begin{IEEEbiography}[{\includegraphics[width=1in,height=1.25in,clip,keepaspectratio]{photo/Wei_Dong}}]{Wei Dong}
	(S’08–M’12) received the B.S. and Ph.D.	degrees from the College of Computer Science, Zhejiang University, in 2005 and 2011, respectively.
	He is currently a Full Professor with the College of Computer Science, Zhejiang University, where he leads the Embedded and Networked Systems Laboratory. 
	He has published over 100 papers in prestigious conferences and journals, including MobiCom, INFOCOM, ToN, and TMC. 
	He is a member of the IEEE and ACM.
\end{IEEEbiography}
\fi



\end{document}